\documentclass[preprint]{aastex}
\include{psfig}
\usepackage{emulateapj5}
       
\newcommand\eg{{\it e.g.} }
\newcommand\etal{et~al.}
\newcommand\ie{{\it i.e.~}\ }
\newcommand\Lya{Ly$\alpha$}

\newcommand\CIVfull{\hbox{C~$\rm IV$}~$\lambda\lambda$~1549}

\newcommand\Ha{H$\alpha$}

\newcommand\OIII{[\hbox{O~$\rm III$}]}
\newcommand\OIIfull{\hbox{[O~$\rm II$]}~$\lambda$~3727}
\newcommand\OIIIfull{\hbox{[O~$\rm III$]}~$\lambda$~5007}
\newcommand\HI{\ion{H}{1}}
\font\aipsfont = cmsy9 scaled\magstep1

\newcommand\aips {{\aipsfont AIPS$\;$}}

\def\spose#1{\hbox to 0pt{#1\hss}}
\def\simlt{\mathrel{\spose{\lower 3pt\hbox{$\mathchar"218$}}
     \raise 2.0pt\hbox{$\mathchar"13C$}}}
\def\simgt{\mathrel{\spose{\lower 3pt\hbox{$\mathchar"218$}}
     \raise 2.0pt\hbox{$\mathchar"13E$}}}

\newcommand\aasup{{A\&AS}}

\slugcomment{To appear in the January 2002 Astronomical Journal}

\shorttitle{Imaging of USS sources}
\shortauthors{De Breuck et al.}

\begin{document}

\title{Optical and Near-IR Imaging of Ultra Steep Spectrum Radio Sources - \\ The $K-z$ diagram of radio and optically selected galaxies. \thanks{Based on observations obtained with the Keck, Lick 3m, CTIO, ESO 3.6m, and William Herschel telescopes.}}
\author{Carlos De Breuck\altaffilmark{2,3,4,5}, Wil van Breugel\altaffilmark{4}, S.\ A.\ Stanford\altaffilmark{4,6}}
\author{Huub R\"ottgering\altaffilmark{5}, George Miley\altaffilmark{5}, Daniel Stern\altaffilmark{7,8}}
\email{debreuck@iap.fr, wil@igpp.ucllnl.org, adam@igpp.ucllnl.org, rottgeri@strw.leidenuniv.nl, miley@strw.leidenuniv.nl, stern@zwolfkinder.jpl.nasa.gov}

\altaffiltext{2}{Institut d'Astrophysique de Paris, 98bis Boulevard Arago, 75014 Paris, France}
\altaffiltext{3}{Marie Curie Fellow}
\altaffiltext{4}{Institute of Geophysics and Planetary Physics, Lawrence Livermore National Laboratory, L$-$413, P.O. Box 808, Livermore, CA 94550, U.S.A.}
\altaffiltext{5}{Leiden Observatory, P.O. Box 9513, 2300 RA Leiden, The Netherlands}
\altaffiltext{6}{Physics Department, University of California, Davis, CA 95616, U.S.A.}
\altaffiltext{7}{Astronomy Department, University of California at Berkeley, CA 94720, U.S.A.}
\altaffiltext{8}{Current address: Jet Propulsion Laboratory, California Institute of Technology, Mail Stop 169-327, Pasadena, CA 91109, U.S.A.}

\begin{abstract}

We present optical and/or near-IR images of 128 ultra steep spectrum (USS)
radio sources. Roughly half of the objects are identified in the optical
images ($R \simlt 24$), while in the near-IR images, $>$94\%
are detected at $K\simlt 22$. The mean $K-$magnitude is $\bar{K}=19.26$
within a 2\arcsec\ diameter aperture. The distribution of $R-K$ colors
indicates that at least 1/3 of the objects observed have very red colors ($R-K>5$). The major axes of the identifications in $K-$band
are preferentially oriented along the radio axes, with half of them having compact morphologies.

The 22 sources with spectroscopic redshifts and $K-$band magnitudes
follow the $K-z$ relation found from previous radio samples, but with a larger
scatter. We argue that this may be due to a dependence
of $K-$magnitude on the radio power, with the highest radio power sources
inhabiting the most massive host galaxies. We present a composite $K-z$
diagram of radio-loud and radio-quiet galaxies, selected from the
HDF-North and the Hawaii surveys. Out to $z \simlt 1$, the radio-loud
galaxies trace the bright envelope of the radio quiet galaxies, while
at $z \simgt 1$, the radio-loud galaxies are $\simgt 2$ magnitudes
brighter. We argue that this is not due to a contribution from the AGN
or emission lines. This difference strongly suggests that radio galaxies pinpoint
the most massive systems out to the highest known redshifts, probably
due to the mutual correlation of the mass of the galaxy and the radio
power on the mass of the central black hole.

\end{abstract}


\section{Introduction}

High redshift radio galaxies (HzRGs) provide some of the best
opportunities to study the spatially resolved emission in galaxies
out to the highest redshifts \citep[$z=5.19$; ][]{wvb99}. Although they
are no longer the only class of galaxies detected at such redshifts
\citep[\eg][]{stei99,spi98,wey98}, they are still amongst the most
massive forming galaxies known. Evidence that HzRGs are indeed massive
galaxies in early phases of their formation is based on several
observations.  First of all, at low redshift ($z\simlt 1$), the host
galaxies of powerful radio sources at are uniquely identified with
massive ellipticals \citep[\eg][]{bes98,mcl00}. Secondly, the tight
correlation in the Hubble $K-z$ diagram for powerful radio sources
suggests that they may also be associated with such galaxies out to $z
\sim 5$ \citep[\eg][]{eal97,wvb98,lac00}. Thirdly, many of the HzRGs 
consist of numerous components \citep{wvb98,pen99} {\it each} of which 
have sizes, UV$_{rest}$ luminosities and star formation rates (SFRs) similar
to the `Lyman Break galaxies' galaxies found in field surveys at
$z \sim 3 - 4$ \citep[\eg][]{stei99}. Fourth, in at least two HzRGs
there is direct, spectroscopic evidence for massive star formation based
on stellar absorption--line spectra \citep{dey97,dey99a}. 
Fifth, redshifted molecular (CO) lines \citep{pap00} and sub-mm \citep{ivi98,arc01,reu01} detections have been reported for many HzRGs:  the inferred SFRs of $\sim 1500~{\rm M}_{\odot} {\rm yr}^{-1}$ are sufficient to form massive galaxies if sustained over $\sim 1$ Gyr.

Because HzRGs are more luminous and larger than field galaxies at similar
redshifts, detailed studies of their optical and near-IR morphologies
can be carried out. For example, deep $K-$band imaging (rest--frame optical) 
has indicated a
morphological evolution in the host galaxies of the most powerful radio
sources: at $z \simgt 3$, they display faint, large-scale morphologies,
often surrounded by multiple components aligned with the radio source,
while at $z \simlt 3$, they appear as a single, compact structure without
radio-aligned components \citep[][hereafter vB98]{wvb98}. Using {\it HST/NICMOS} $H-$band data, \citet{pen01} found a similar change from objects with de Vaucouleurs profiles at $z \le 2.3$ to irregular at higher redshifts. 
In the rest-frame UV, this alignment is even more pronounced, and it occurs
in most of $z\simgt 0.7$ powerful radio galaxies \citep[\eg][]{cha87}. A
variety of physical mechanisms has been proposed to explain the nature
of this alignment effect. They include (i) star-formation
induced by shocks associated with the radio jet propagation outward from
the central AGN \citep[\eg][]{dey97,bic00}, (ii) scattering of light from
an obscured nucleus by dust or electrons \citep[\eg][]{cim93,ver01},
and (iii) nebular continuum emission from warm line emitting clouds
\citep[\eg][]{dic95}. High-resolution optical and near-IR {\it HST}
observations of HzRGs indicate that the aligned light is probably a
mixture of these proposed components \citep{pen99,pen01,ste99b}.

To examine the influence of radio sources on the galaxy formation
process, it is important to obtain a large sample of HzRGs, covering a
range in redshift and radio power.  Despite several intensive search
campaigns during the last two decades (see Table \ref{surveys}), the number of known HzRGs is
still quite limited: at present only 26 radio galaxies are known with
$z>3$. Furthermore, because many of the HzRGs, at least at $z < 3$, were
discovered from flux limited radio samples, there is a strong 
dependence of radio power on redshift in the samples of known HzRGs.
We therefore constructed a new sample whose aims are (i) to substantially increase the sample of 
known $z>3$ HzRGs, and (ii) to obtain a significant number of $2<z<3$ 
radio galaxies with radio powers an order of magnitude less 
than the bright samples.
We therefore constructed a sample of ultra steep spectrum (USS, $\alpha <
-1.30; S \propto \nu^{\alpha}$) radio sources drawn from new radio surveys
that reach flux density levels more than an order of magnitude fainter
than previous surveys. USS sources have been successfully used for over
two decades to find HzRGs \citep[\eg][]{rot94,cha96,jar01a}. This technique mainly makes use
of a radio $k-$correction effect: at higher redshift, an increasingly
steeper part of the generally concave radio spectrum shifts to the fixed
observing frequencies of the large radio surveys.

This paper is the second in a series of three describing the observations
of our sample of 669 USS sources. In Paper~I \citep{deb00a}, we defined
our sample. We also presented VLA and ATCA radio images of 410 sources,
which provide radio morphological information and the accurate positions
needed to identify the optical and/or near-IR counterparts. In this
paper, we present the optical and/or near-IR imaging of a sub-sample
of 128 sources. We shall use the near-IR magnitudes to obtain redshift
estimates based on the Hubble $K-z$ diagram. In Paper~III \citep{deb01},
we present optical spectroscopy of 46 sources from this imaged sub-sample,
including 34 sources with spectroscopic redshifts.

The layout of this paper is as follows. In \S2, we describe the optical
and near-IR observations and data reduction, paying special attention
to the astrometry. We present the results, discuss the morphologies
of the near-IR identifications and examine the correlations between
the various parameters in \S3. In \S4, we concentrate on the Hubble
$K-z$ diagram.  We discuss the expected redshift distribution from our
sample in \S5. We present our conclusions in \S6. Throughout, we assume
$H_0=65~$km~s$^{-1}$~Mpc$^{-1}$, $\Omega_M=0.3$, and $\Omega_{\Lambda}=0$.

\section{Observations and Data Reduction}

We obtained the optical and near-IR images described in this paper at five different observatories between 1996 April and 2000 January.  
Table \ref{imobservations} gives an overview of the 21 successful imaging observing sessions. We use a consecutive numbering scheme for each session in the project, counting also the sessions where we did not obtain any data, or only spectroscopic data (see Paper~III\nocite{deb01}). Table \ref{imjournal} gives a journal of the observations, and Table \ref{imsetup} gives an overview of the instrumental setups.

\subsection{Target selection}
We selected all our targets from the USS sample in paper~I. Although we selected the objects at each individual session on the basis of the accessible right ascension range, we have tried to distribute the objects over a large range of spectral index, radio morphology, and 1.4~GHz flux density. The most important improvement of our sample with respect to other recent surveys is the lower flux density limit: the median flux density of our faintest sample (WN) is almost an order of magnitude below all other efforts (Table \ref{surveys}). We therefore selected half of the objects observed in $K-$band with the additional criterion $S_{1400} < 50$~mJy. Although this paper presents only the first batch of images ($\sim$13\% of the sample in $R$ and $K$), we expect it to be fairly representative for the sample as a whole.

\subsection{Optical Imaging}

\subsubsection{Observations}

\paragraph{Lick 3m}

We used the Kast imaging spectrograph \citep{mil94} at the Lick Observatory Shane 3m telescope on Mount Hamilton to obtain images in the ``Spinrad night-sky'' filter. This yields red magnitudes in the $r_S$ system, which has an effective wavelength of 6890~\AA, and an equivalent width of 1370~\AA; the $r_S$ magnitudes can be related to commonly used photometric systems as determined by \citet{djo85}\footnote{The transformation from Spinrad night-sky $r_s$ into Johnson $V R$ is $(r_S-R)=-0.004-0.072(V-R)+0.073(V-R)^2$.}. The field of view of Kast is $\sim$145\arcsec\ square, but there is substantial circular vignetting in all corners. 
To enlarge the field of view and reduce the chance that the object falls on a bad pixel on the CCD, we dithered the individual exposures by $\sim$15\arcsec. We limited the exposure times of the individual exposures to 3$-$5~minutes to avoid extensive saturation of bright stars in the field.

\paragraph{ESO 3.6m}

To identify the radio source for spectroscopic observations, we took 10$-$15~minute images with the ESO faint object spectrograph and camera \citep[EFOSC1;][]{sav97} on the ESO 3.6m telescope at La Silla. If an object was not detected in 15~min, we did not attempt to obtain a spectrum. We re-observed a large number of these optical ``blank fields'' in $K-$ band at Keck observatory (see \S2.3.1).

\paragraph{WHT}

On two occasions, we used the prime focus imaging platform \citep[PFIP;][]{car95} at the William Herschel Telescope at La Palma to obtain deep $R-$band images of USS sources. During the 1997 May run, we used the LORAL CCD, while during the 1999 May run, we used the EEV12 CCD. This resulted in slightly different pixel scales and fields of view (see Table \ref{imsetup}).

\paragraph{Keck}

We obtained $I-$band images of 3 USS sources with the Low Resolution Imaging Spectrometer \citep[LRIS;][]{oke95} on the Keck~II telescope on Mauna Kea. The main use of these deep images was to determine accurate positions and offsets from nearby brighter stars ($I \simlt 18$) to set up the spectroscopic observations during the same observing session.

\subsubsection{Data Reduction}

For all the imaging data, we performed standard data reductions consisting of overscan bias correction, and flat fielding using a median sky flat, constructed for each individual night from the non-saturated science exposures. 

To correct for the circular edges of the Lick images before registering them, we constructed a filter image from the flat-field frame, filtering out all pixels at $\simlt$96\% of the median central flat-field value. Failure to do so would lead to a strongly increased noise pattern in the outer regions of the registered image.
To register the dithered Lick images, we first determined the relative position-shifts between the images using the task {\tt registar} from the IRAF DIMSUM\footnote{DIMSUM is the Deep Infrared Mosaicing Software package, developed by P. Eisenhardt, M. Dickinson, A. Stanford, and J. Ward, which is available as a contributed package in IRAF.} package. Finally, we combined all individual image frames using integer pixels shifts. This is possible, because the PSF is over-sampled by a factor of 2 to 3 (depending on seeing). The use of fractional pixel shifts would require re-sampling of each individual image, which would smear out the PSF in the combined image.

The astrometric calibration and identification of the host galaxy is described in \S 2.2.3.

We flux calibrated the images using observations from several standard stars obtained during the same night. After including the zero-point, aperture centering and sky background fitting uncertainties given by the {\tt phot} routine, we estimate the photometric uncertainty to be $\sim$0.2 magnitudes. We used the IRAF task {\tt phot} to determine fixed aperture magnitudes. We used a 4\arcsec\ diameter aperture for the optical images, because the seeing was often not sufficiently good to justify using a smaller aperture. For images obtained under non-photometric conditions, we used a value of the photometric zero-point determined from previous observations; these magnitudes (indicated with $\dag$ in Table \ref{rphotometry}) should be considered as indicative only.

\subsection{Near-IR Imaging}
\subsubsection{Observations}

We used near-IR cameras at the Keck and CTIO observatories to observe sources in both the northern and southern hemispheres. At each observatory, we employed a non-redundant 16$-$point dithering pattern. The integration times and specific $K-$band filter ($K, K_S$ or $K^{\prime}$) were adapted to the expected magnitudes and sky conditions. We first describe the site-specific details, and next describe the data reduction procedure.

\paragraph{Keck}

The objects that remained undetected after optical imaging ($R \simgt 24$) constitute excellent candidates for the highest redshift radio galaxies. We observed these objects with the near-infrared camera \citep[NIRC;][]{mat94} on the Keck~I telescope on Mauna Kea. NIRC contains a $256\times 256$ InSb array with a pixel scale of 0\farcs151 pixel$^{-1}$. Depending upon the sky conditions (as measured from the sky counts during the observations), we used either a standard $K$, a $K_S$ or a $K^{\prime}$ filter. We used typical integration times of 60~s per pointing, comprised of 3 co-added 20~s frames, in a 16-point non-redundant dithering pattern. Where possible, we reduced the images of the first 16-point cycle during the integration on the next cycle. If the object was clearly detected in one cycle, we did not start a third cycle, in order to save observing time for other sources.

The NIRC requires a $R \simlt 18$ guide star to be placed in an offset guider. To achieve this, we often needed to rotate the target field. In some cases, we also had to shift the field by several arcseconds in order to place the guide star on the guider camera. As a result, the images obtained have to be rotated back to orient them in the standard North up, East left orientation, and the objects are not always perfectly centered in the registered frames.
Bright stars in the frames display electronic 'bleed trails', for which we do not attempt to correct in the final images. 

The data from the 1999 September observing session (K12) were severely affected by instrumental problems due to accidental external illumination of one corner of the detector. This decreased the gain down to $<$10\% of the normal value over more than a third of the detector array. During this session, we adapted our dithering pattern such as to avoid placing the identification in this affected areas. These areas are clearly visible in the finding charts shown in the Appendix. The measured counts of the objects in the relevant corners are severely attenuated by this problem, a factor that should be taken into account if the resulting images are being used as finding charts.

\paragraph{CTIO}

For the southernmost sources from our USS sample, we used the CIRIM camera with the tip-tilt system on the CTIO Blanco 4m telescope at Cerro Tololo in the f/14 focal ratio. The detector is a $256\times 256$ HgCdTe NICMOS 3 array. We used typical integration times of 120~s per pointing, comprised of 6 co-averaged 20~s frames, in a 16-point non-redundant dithering pattern. To improve the resolution in the images, we performed a tip-tilt correction, using a $V<16$ reference star within 2\farcm5 of the radio galaxy.

\subsubsection{Data Reduction}

We reduced the near-IR data using the NOAO IRAF package. After bias subtraction and flat-fielding, we sky-subtracted, registered, and summed the data using the DIMSUM near-IR data reduction package. For most images observed with excellent seeing ($\simlt$0\farcs5), we block-replicated the pixels by a factor of 2 before summing the individual images.

\subsubsection{Astrometry}

The mean density of objects down to $R=25$ from faint galaxy counts is $\sim 10^{5}$~deg$^{-1}$~mag$^{-1}$ \citep{sma95}. This agrees well with the space density of $R\simlt$24.75 objects in the field of USS radio sources found by \citet{rot95}. Because the optical images reach similar or deeper limiting magnitudes, we can statistically expect to find an unrelated source within $\sim$11\arcsec\ from the radio position. This can lead to a number of possible mis-identifications in cases where the unrelated serendipitous source falls within $\simlt$2\arcsec\ from the radio position. One way to limit the number of mis-identifications is to use more accurate astrometry. The accuracy of an astrometric solution is dominated by 2 factors: (i) the positional uncertainty of the astrometric reference catalog with respect to the International Celestial Reference Frame (ICRF), and (ii) the accuracy of the transformation of this astrometrical information to the images.
The uncertainty in the transformation is mainly influenced by the number of objects available to solve the astrometric parameters (positional zero-points in the x- and y-directions, pixel scales, and rotation) using the positions from the astrometric reference catalog. Because the field of view of the images is generally only a few square arcminutes and we are mainly interested in the central $\sim$arcmin$^2$, we did not attempt to determine the field distortion parameters. 

The uncertainties of the VLA positions with respect to the ICRF are $\simlt$0\farcs2. As we will show below, this is 2 to 3 times smaller than the uncertainty between the optical astrometry and the ICRF. In the following, therefore, we will consider only the uncertainties in the optical images. Note that this does not include the uncertainty in the predicted position of the optical/near-IR identification with respect to the radio morphology. In cases where the radio source consists of different components it is possible that the identification does not fall at the midpoint of the radio lobes, but closer to one of the lobes \citep[see \eg][]{deb99}. In those cases, a visual inspection of the radio source overlaid on the optical/near-IR image often provided an unambiguous identification.

For the optical and near-IR images obtained prior to 1998, we used the astrometric solution from the digitized sky survey (DSS-I). \citet{ver96} compared accurate absolute VLBI positions of 153 QSO's with positions from the DSS-I, and found the rms uncertainty of the DSS-I astrometry is $\sim 0\farcs6$ in both coordinates over the whole sky. 
In 1998, the more accurate USNO-A2.0 catalog was published \citep{mon98}, which is based on astrometry from the Tycho catalog \citep{per97}, generated from the Hipparcos mission. The astrometric uncertainty of this catalog with respect to the ICRF is more than 2 times better: $1\sigma \approx 0\farcs25$ \citep{deu99}. Provided that a good transformation of this astrometry to the object frames is possible, we can identify the radio source to within one seeing element in the optical and near-IR images. 

To perform the astrometric calibration on the images, we used the task {\tt XTRAN} in the NRAO \aips\ package. In almost all cases, the optical images were obtained before the near-IR image, so we can use the larger field of view of the optical images (see Table \ref{imsetup}) to identify the field with the DSS images.

For the astrometric calibration, we only consider stellar objects in the field, because the inclusion of resolved galaxies would introduce additional uncertainties to our solution. In some cases, we had to remove some of the reference objects from our solution in order to get a better fit; these objects are most likely stars with a significant proper motion. The mean number of available reference stars in the optical images is 8 (median is 7). Table \ref{imsetup} lists the mean pixel scales and their $1\sigma$ deviations, as derived from the {\tt XTRAN} solutions. We find that the uncertainties in the optical pixel scales are generally $\simlt$0\farcs001/pixel, with the largest uncertainties occurring in the images with the smallest field of view, viz.\ those obtained with Kast at Lick.

In frames where less than 4 reference stars were available, instead of following the above procedure, we used the position of a single star as the origin of the coordinate system, and adopted the pixel scale quoted in Table \ref{imsetup}. When a second or third reference star was available, we selected the object that provided the most consistent results, or the object that was closest to the image center. 

Because the field of view of the registered $K-$band images barely exceeds 1\arcmin$\times$1\arcmin, we rarely have sufficient stars from the DSS-I, DSS-II or the USNO-A2.0 catalog to perform the same astrometric calibration as in the optical images. We therefore translated the astrometry from the optical images, using stellar objects to $R\simlt 24$ in common between the optical and near-IR frames. On average, we used 7 common objects to solve the astrometry in the NIRC images. For the $K-$band images obtained at CTIO, no optical images deeper than $R \sim 20$ were available, and the uncertainties are significantly larger. Nevertheless 6 radio galaxy identifications determined using this astrometry have been shown to be genuine, as they were all successfully detected spectroscopically with the VLT (see Paper~III).

For 4 objects having $K-$band images, but lacking large field of view optical images to find reference objects, we obtained $J-$ and $K-$band images using the Gemini twin-arrays infrared camera \citep{mcl93} at the Lick 3m telescope. In addition, R. Gal kindly obtained wide field $r-$band images of 3 additional objects (TN~J2008$-$1344, TN~J2028$-$1934, WN~J2221$+$3800) with the Palomar 60\arcsec\ telescope. We translated the astrometric solution obtained with these shallow wide-field images to align the deeper, but smaller $K-$band images.

Using the astrometric solutions, we could find unambiguous identifications within $\simlt$1\arcsec\ of the radio positions in $>$90\% of the optical images, and $>94$\% of the near-IR images (considering only the images with a possible identification within 8\arcsec\ from the radio position). In cases where there was another object within $\simlt$2\arcsec, we adopted as the real identification the object that was closest to the line connecting both radio lobes, or the object with the more diffuse morphology (assuming the other unresolved object to be a foreground star).

\subsubsection{Photometry}

We calibrated the photometry using short observations of standard stars from the UKIRT faint standard list \citep{cas92} or from the list of \citet{per98}. For NIRC, this procedure yielded typical zeropoints of $K_0=25.12 \pm 0.03$, ${K_S}_0=24.75 \pm 0.03$, and $K^{\prime}_0=25.12 \pm 0.03$ (for 1 count/second, integrated over the source). For CIRIM at the CTIO~4m, we found $K_0=21.73 \pm 0.05$. We estimated the uncertainties from the variation of the zeropoints determined from different standards during the same night. The zero-points include the airmass term, but we did not correct for airmass variations between the targets, because we observed all of our objects with airmasses $<$1.7 (mean airmass 1.15), and the airmass dependence in $K-$band is small compared to the aperture centering and sky fitting uncertainties described below.

Because in most cases the redshift was not known at the time of the near-IR observations, we could not predict the expected $K-$band magnitudes, as determined from the Hubble $K-z$ diagram (\S 4). Consequently, there is a significant range in signal to noise (S/N) in the fixed apertures of the final images, with some of the brighter images reaching S/N levels well over 50, while the faintest detections are only detected down to the 5$\sigma$ level. As a result, the use of isophotal magnitudes is inappropriate. To determine a standard fixed aperture, we first measured the magnitude in a variety of different apertures, with diameters ranging from 0\farcs4 to 8\arcsec, using the IRAF task {\tt phot}. Figure \ref{Kprofiles} shows the resulting variation of the aperture magnitudes as a function of the aperture diameter. It is clear that for the large majority of sources, most of the flux is contained within a 2\arcsec\ diameter aperture. This is also visible from the images (Fig. \ref{Kcutouts}).

Table \ref{kphotometry} therefore lists magnitudes in 3 different apertures: (i) a 2\arcsec\ aperture, which, as shown above, corresponds best to the actual size of most galaxies, (ii) a 4\arcsec\ aperture for comparison with the optical magnitudes in Table \ref{rphotometry}, and (iii) an 8\arcsec\ aperture for comparison with previously published $K-$band photometry in the literature \citep[\eg][vB98]{eal97}.  The uncertainties quoted are our best estimates, which include both the zero-point uncertainty and the aperture centering and sky fitting uncertainties given by the {\tt phot} routine.

We did not correct the magnitudes for Galactic extinction, but list the expected extinction in the $K-$band in Table \ref{kphotometry}. We obtained these values using the NASA Extragalactic Database (NED), which are based on the $E(B-V)$ values from the extinction maps of \citet{schl98}, and converted to $A(K)$ using the extinction curve of \citet{car89}, under the assumption of $R_V=A(V)/E(B-V)=3.1$. The $A(K)$ corrections (listed in Table \ref{kphotometry}) are generally negligible compared to the uncertainties, except in a few objects which lie close to the Galactic plane.

\section{Results}

Figure \ref{Kcutouts} shows $8\arcsec \times 8\arcsec$ images of the $K-$band images, centered on the derived identifications (\ie\ not the predicted radio position). The grey scale is from $\mu - \sigma$ to $\mu + 4 \sigma$, where $\mu$ and $\sigma$ are the mean and rms, as determined from a histogram of the sky-subtracted background counts, determined in a source-free region with the \aips task {\tt IMEAN}. We show $90\arcsec \times 90\arcsec$ finding charts in the Appendix. We show only one finding chart per object, selecting the near-IR image where available, because it is of higher quality than the optical images. All images are shown in the usual orientation with North up and East to the left. Note that this required a generally non-orthogonal rotation of the NIRC images because they were obtained in a orientation that positions a guide star in the offset guider camera (see \S2.3.1).
We shall now discuss the optical and/or near-IR identification of a number of individual sources, and then consider the magnitude distribution of the sample as a whole. $K-$band observations of 2 sources (WN~J0117$+$3715 and WN~J1843$+$5932) from our USS sample have been reported by \citet{vil99}.

\subsection{Notes on individual sources}

{\bf WN~J0034$+$4142}: Although the near-IR identification does not lie at the midpoint, it falls on the line connecting both radio lobes.

{\bf WN~J0117$+$3715}: \citet{vil99} report $K=18.20 \pm 0.57$ in a 3\farcs88 diameter aperture for this source, which is consistent with our value ($K=17.57\pm0.04$ in a 4\arcsec\ diameter aperture).

{\bf MP~J0130$-$8352}: No scale size could be estimated for this source because there is no appropriate PSF-star in the CTIO image.

{\bf MP~J0202$-$5425}: There is a bright stellar object $\sim$3\arcsec\ to the north-east of the near-IR identification, which contaminates the aperture photometry, and the determination of the spatial profile of this source.

{\bf TN~J0205$+$2242}: This galaxy has been spectroscopically confirmed at $z=3.506$. The near-IR surface brightness profile appears very steep for an object at such a redshift, and resembles that of a radio galaxies at $z<3$.

{\bf TN~J0218$+$0844}: The most likely optical identification is the source slightly north of the predicted radio position, although we cannot exclude the source at $\alpha_{J2000}=2^h18^m25\fs57, \delta_{J2000}=08\arcdeg44\arcmin27\farcs7$. Because we did not observe this source in $K-$band, we do not show a cutout image of this source in Fig.~\ref{Kcutouts}.

{\bf MP~J0249$-$4145}: No sources are detected down to $K\sim 19$ within the positional uncertainty of our ATCA radio map.

{\bf WN~J0305$+$3525}: The near-IR identification is an very diffuse source without a central core component.

{\bf WN~J0346$+$3039}: The near-IR photometry and morphology for this object is highly uncertain due to the problems with the detector during this run.

{\bf WN~J0359$+$3000}: The near-IR identification is uncertain because the radio position is based on the NVSS position only \citep[which has an rms position uncertainty of $\sim 1\farcs5$ for this source;][]{con98}. The near-IR photometry and morphology is also highly uncertain due to the problems with the detector during this run.

{\bf MP~J0449$-$5449}: No scale size could be estimated for this source because there is no appropriate PSF-star in the CTIO image.

{\bf WN~J0617$+$5012}: The very faint $K-$band identification coincides with the southern radio lobe of a $z=3.153$ radio galaxy (see Paper~III).

{\bf WN~J0741$+$5611}: Our radio map is of insufficient depth to identify any of the features observed in the field of the $K-$band image. We therefore do not show a cutout image of this source in Fig.~\ref{Kcutouts}.

{\bf TN~J0856$-$1510}: Our radio map is of insufficient depth to determine the exact optical or near-IR identification. There are 2 candidate identifications in the $K-$band image, one at $\alpha_{J2000}=08^h56^m12\fs40, \delta_{J2000}=-15\arcdeg10\arcmin40\farcs4$, and one at $\alpha_{J2000}=08^h56^m12\fs37, \delta_{J2000}=-15\arcdeg10\arcmin32\farcs7$. We therefore do not show a cutout image of this source in Fig.~\ref{Kcutouts}.

{\bf TN~J0924$-$2201}: This is the radio galaxy with highest known redshift \citep[$z=5.19$;][]{wvb99}.

{\bf TN~J0936$-$2243}: The central component and the companion object $\sim$4\arcsec\ to the north both have $z=1.479$. This source is discussed in detail in Paper~III.

{\bf WN~J1015$+$3038}: The 3 companion objects may well be part of the same physical source, which is probably at $z=0.54$ (see Paper~III). If this is the correct, this is object is more than a magnitude redder in $R-K$ than the other radio galaxies at these redshifts (Fig. \ref{zRmK}).

{\bf TN~J1033$-$1339}: No scale size could be estimated for this source because there is no appropriate PSF-star in the NIRC image.

{\bf TN~J1049$-$1258}: The string of objects is aligned with the radio source, which also consists of multiple knots.

{\bf WN~J1053$+$5424}: The near-IR identification is a very diffuse source without a central core component, even though the radio source is a small 1\farcs1 double (see paper~I).

{\bf WN~J1123$+$3141}: The near-IR identification of this large radio source (largest angular size $25\farcs8$) at $z=3.217$ is a spectacular interacting system showing extended tidal tails. The $K-$band magnitude is $\sim$2 magnitudes brighter than the main trend (Fig. \ref{Kzradio}).

{\bf WN~J1314$+$3649}: This is one of our faintest $K-$band identifications. We attempted optical spectroscopy with LRIS at Keck, but did not detect the object (see Paper~III).

{\bf TN~J1338$-$1942}: The near-IR identification, a radio galaxy at $z=4.11$, corresponds to the brightest component of a very asymmetric radio source. This source is described in detail by \citet{deb99}.

{\bf WN~J1355$+$3848}: No scale size could be estimated for this source because there is no appropriate PSF-star in the NIRC image.

{\bf TN~J1428$+$2425}: This is a rare $K-$band ``blank field''. Although our astrometry is based on only 2 stars in common with an image taken with the Gemini camera at Lick observatory, there are no candidate identifications within 8\arcsec. We therefore do not show a cutout image of this source in Fig.~\ref{Kcutouts}.

{\bf WN~J1525$+$3010}: There appears to be a diffuse halo around the $K-$band identification. We have obtained a $t_{int}=5400$~s spectrum with LRIS at Keck, but did not detect any optical emission from this source (see Paper~III).

{\bf WN~J1550$+$3830}: No scale size could be estimated for this source because there is no appropriate PSF-star in the NIRC image.

{\bf WN~J1604$+$5505}: The near-IR identification is a very diffuse source without a central core component.

{\bf WN~J1731$+$4654}: This is a rare $K-$band ``blank field''. We obtained the astrometric solution by transferring a single star from the Lick image and assuming the pixel scale; comparison with 2 other stars common with the Lick image suggests this solution is accurate up to $\sim$1\arcsec. There is a source 5\arcsec\ north of the predicted position at $\alpha_{J2000}=17^h31^m59\fs68, \delta_{J2000}=46\arcdeg54\arcmin05\farcs5$, but this falls well outside the position uncertainty from the FIRST survey. We therefore do not show a cutout image of this source in Fig.~\ref{Kcutouts}.

{\bf WN~J1836$+$5210}: The optical and near-IR identifications are ambiguous between a red northern component, and a blue southern component. Table \ref{astrometry} lists the position of the red, northern component, which is slightly closer to the predicted radio position, although both sources might well be part of the same physical system.

{\bf WN~J1843$+$5932}: \citet{vil99} detected this source with $K=17.96 \pm 0.19$ in a 3\farcs28 aperture. We did not re-observe it, and refer to their paper for a finding chart.

{\bf TN~J2009$-$3040}: This source has a strong unresolved component, probably a direct contribution from the AGN.

{\bf TN~J1941$-$1952}: The unresolved object 1\arcsec\ west of the near-IR identification is probably a foreground star, and contaminates the aperture photometry.

{\bf WN~J2044$+$7044}: This object appears compact in the $K-$band image. We attempted optical spectroscopy with LRIS at Keck, but did not detect any optical emission from the object (see Paper~III).

{\bf WN~J2213$+$3411}: The near-IR identification is uncertain, because we lack a high resolution VLA map of this field. However, the NVSS map shows a source that is elongated in the same direction as the proposed identification ($52\arcsec \times 19\arcsec$ at position angle $-$31\arcdeg).

{\bf WN~J2313$+$4253}: This is the pulsar PSR~J2313+4253 \citep{kap98,han99}.

\subsection{Magnitude and color distributions}

We identified approximately half of the 83 optically imaged USS sources, while in the near-IR, we found counterparts for $>$94\% of the 86 sources observed. Of the 4 $K-$band non-detections, MP~J0249$-$4145 was obtained with CIRIM at CTIO, and might well have been detected if observed down to fainter levels comparable to the Keck data. Another non-detection, WN~J0741$+$5611, requires more sensitive radio map to determine the possible identification. The remaining 2 sources, TN~J1428$+$2425 and WN~J1731$+$4654, have good radio maps (see paper~I), and could be galaxies with extremely faint $K-$band identifications or Galactic pulsars.
 
It is clear that even a moderately deep ($K \simlt 20$) near-IR survey of radio sources with a 4m class telescope is much more efficient in detecting the galaxy counterparts than deep ($R \simlt 25$) optical imaging. Once identified, the offsets from brighter stars in the field of the radio galaxy allow one to accurately position the slit for spectroscopic redshift determinations. Moreover, the $K-$band magnitude provides an estimate of the expected redshift of the radio galaxy by means of the Hubble $K-z$ diagram (\S 4).

Because the magnitude limits of our optical imaging vary from $R\sim23$ to $R\sim26$, and half of the observed USS sources are not detected, we cannot construct a reliable optical magnitude distribution. Only one third of the sources are brighter than $R=24$, suggesting that the mean optical magnitude is probably $R\sim 25$. This is significantly fainter than that of the USS sample of \citet{rot95}, who found that the $R-$band magnitude distribution in their sample peaks between $R=22$ and $R=23$, with 70\% of the radio sources identified at $R<24$. This discrepancy is due to 2 main differences between the samples. First, the flatter radio spectral index cutoff ($\alpha \sim -1.2$) in their sample, which selects lower redshift objects than our optically observed sample, with a mean (and median) spectral index of $\bar{\alpha} \simeq -1.38$. Second, their sample contains sources with radio flux densities an order of magnitude brighter than our sample.

Our optical magnitude distribution more resembles that of the sample of fainter USS sources by \citet{wie92}, who find that half of their sources have $R>24$. The \citet{wie92} sample has a similar spectral index selection ($\alpha_{327}^{608} < -1.1$) as the \citet{rot95} sample, but has flux densities comparable to our WN sample. This suggests that the lower radio flux densities may explain the fainter optical magnitudes, because the host galaxy absolute magnitude appears to be correlated with radio luminosity \cite[\eg][]{bes98,lac00}. Such a correlation could arise if the host galaxies of the more powerful radio sources contain a greater mass of stars, and assuming that their central black holes accrete at near-Eddington rates. We return to this point in \S 4.2.

Figure \ref{Khis} shows the distribution of $K$ magnitudes. The mean/median $K-$band magnitude is $K=19.26/19.19$ in a 2\arcsec\ diameter aperture ($K=18.57/18.72$ in a 4\arcsec\ aperture). The distribution appears to be slightly asymmetric, in the sense that the tail of sources with magnitudes fainter than the mean seems to be broader than the bright-end tail, which cuts off at $K=17.5$. This may be partially due to our prior selection of objects undetected on the POSS ($R \simlt 20$).

We can get an idea of the content of the sources in our sample from the $R-K$ color distribution shown in Figure \ref{R-Khis}. This histogram consists of only 37 sources, of which 27 are non-detections in the $R-$band. This is mostly due to the nature of our observational program, where we assigned lower priorities in our $K-$band imaging campaign to those objects that had been previously detected in the $R-$band. From this histogram we can conclude that at least half of the sources observed have $R-K > 4$, at least a third have $R-K > 5$, and at least 10\% are bona fide extremely red objects \citep[EROs; \eg][]{els88,gra96}. Note that due to the shallowness of the optical imaging, these are mostly lower limits. We discuss the $R-K$ color evolution in \S 4.3.2.

\subsection{Morphology}

To obtain basic morphological information of the host galaxies (ellipticity and position angle), we used the IRAF$-$STSDAS task {\tt ellipse} to fit surface brightness profiles in the near-IR images. The profiles are generated by fitting elliptical isophotes to the data, with the center, ellipticity, and position angle of each ellipse allowed to vary. The iterative method used in the fitting of each ellipse is described in \citet{jed87}. We fit each source interactively, and continued fitting outward isophotes until the ellipticity or position angle started showing large variations or nearby companions were picked up.
We did not attempt to fit radial profiles for (i) sources observed with CIRIM at CTIO, because the seeing was too poor, (ii) sources with companion objects within $<$2\arcsec, (iii) sources fainter than $K\sim 20$ in a 2\arcsec\ diameter aperture, and (iv) sources with an obvious non-elliptical morphology (\eg TN~J1049$-$1258 and WN~J1123$+$3141). Excluding these sources, we retain 44 objects. Table~\ref{kphotometry} lists their ellipticities and position angles.

A visual inspection of Figure~\ref{Kcutouts} shows the generally compact structure of the $K-$band identifications, even with the good seeing of the Keck images. This is partly due to our observing strategy which cut short the integration times of objects that could be detected in quick mountain reductions of the first frames. Consequently, we did not obtain images with a high signal/noise ratio out to large distances. This is illustrated by the fact that more than half of the flux of the 8\arcsec\ diameter apertures is enclosed within a 2\arcsec\ aperture (See also Fig.~\ref{Kprofiles}). This makes it impossible to determine surface brightness profiles out to sufficiently large radii to differentiate between standard galaxy functional forms such as de Vaucouleurs or power-law disc models.

However, it is possible to estimate the half light radii of the host galaxies, using a procedure described by \citet{lac00}. This method assumes that the host galaxies have de Vaucouleurs profiles \citep[][note that disc models do not give significantly different best-fitting half light radii]{roc98}. To estimate the half light radius in an image, we constructed a set of model galaxies with circular de Vaucouleurs profiles with a range of half light radii. We then convolved these model galaxies with a PSF profile from a nearby star, and normalized them to match the total flux of the observed galaxy. We subsequently subtracted the model galaxies from the data and measured the rms in the residual image. Finally, we fitted a second-order polynomial to the three model galaxies yielding the lowest rms residuals, and determined the half light radius from the minimum of this polynomial.

The derived values need to be corrected for ellipticity. To correct for this, we did not use the ellipticities quoted in Table~\ref{kphotometry}, because the values are derived from the outer isophote only, which may be subject to large uncertainties due to the inclusion of unresolved companion objects. Instead, we assumed a mean ellipticity of $e=0.2$, following \citet{lac00}. Table~\ref{kphotometry} lists the corrected estimates of the half light radii for 36 galaxies where this procedure was successful. The other objects either have no star in the field that can be used as a PSF, have too low signal/noise, or obvious non-elliptical morphologies due to strong interactions with the radio source or companion objects.

The median $r_{hlr}$ is 0\farcs48 (mean is 0\farcs60). This is similar to the values measured by \citet{roc98} for a sample of 6C galaxies at $z\sim 1$ (mean 0\farcs55, median 0\farcs48). However, our values are slightly smaller than those found by \citet{lac00} for a sample of 7C sources (mean 1\farcs24, median 0\farcs90). These authors argue that the smaller $r_{hlr}$ in the 6C sample could have been underestimated due to poor seeing, but our sample was observed with better seeing that the 7C, and still finds smaller $r_{hlr}$. To examine the dependence of $r_{hlr}$ on intrinsic source parameters, such as redshift or radio power, it will therefore be important to obtain more spectroscopic redshifts for our sample (at present, we only have spectroscopic redshifts of 7 objects with measured $r_{hlr}$). We conclude that the present data are insufficient to examine the evolution of the host galaxies of powerful radio sources.

\subsection{Correlations with source parameters}

Because we detected virtually all the sources from our sample we observed in $K-$band, we can now examine correlations between the parameters determined from the near-IR images (magnitude, color, ellipticity, position angle) and other source parameters (redshift, radio power, radio spectral index, radio size, radio position angle, emission line luminosity). Paper~I and III list the radio and spectroscopic parameters, and how they were determined.
Because the vB98 sample consists of well-studied objects, we added their data to our sample to search for possible correlations.

\subsubsection{Correlations with redshift}

The most obvious correlation is between redshift and $K-$band magnitude, the so called Hubble $K-z$ diagram, which is discussed separately in \S 4. In \S 4.3.2, we supplement our sample with data from the literature to study the color evolution of the radio galaxy hosts.

\subsubsection{Correlations between morphological parameters}

The radio and near-IR morphologies do not show strong correlations, except for the tendency for the main axis of the near-IR identification and the radio position angle to be aligned. We return to this point in \S 3.5. 

\subsubsection{Correlations with emission line luminosities}

For less than 20 sources we have information on the emission line luminosity of \Lya\ and/or \CIVfull. Neither these luminosities nor their ratio appear to be correlated with any of the photometric or morphological near-IR parameters, although a weak correlation between these parameters would not be detected in such a small sample.

\subsubsection{$K-$band magnitude and radio spectral index}

Because we use both the steepness of the radio spectral index and the faintness of the $K-$band magnitude to select the highest redshift radio galaxies, it is of interest to examine if both parameters are correlated through their mutual redshift dependence. In our inhomogeneous sample, the low frequency spectral indices and 2\arcsec\ or 4\arcsec\ diameter aperture $K-$band magnitudes are correlated at the 88\% significance level. However, we do not consider this number representative because of (i) the limited spectral index coverage $\alpha < -1.2$ in our sample, (ii) the incomplete $K-$band imaging of $<$13\% of our sample, and (iii) the radio power dependence within our sample (see \S 4). 
To illustrate the importance of this third point, we note that mean/median $K-$band magnitude of sources observed from our radio brightest MP sub-sample (see Paper~I) is $\bar{K}_{MP}=18.6$, while for our $\sim$10 times fainter WN sample, $\bar{K}_{WN}=19.4$. For these same sub-samples, the median spectral indices are $\bar{\alpha}_{MP}=-1.23$ and $\bar{\alpha}_{WN}=-1.41$.

\subsubsection{$K-$band magnitude and radio size}

Figure \ref{Klas} shows the radio size plotted against $K-$band magnitude. The faint $K>20$ sources are all very small in the radio ($\simlt 3\arcsec$), while the brighter sources have a large range of radio sizes. Because the $K-$band imaging has an almost complete identification fraction, and the identifications of the large radio sources are unambiguous, we can exclude that the distribution in Figure \ref{Klas} is due to selection effects.

The spectroscopic observations (see Paper~III) concentrated on these weak sources with small radio sizes. Of the 18 $K>20$ detections 13 were observed spectroscopically. We could determine the redshift of 7 objects, which have a mean of $\bar{z}=3.67$, range from $z=2.54$ to $z=5.19$, and include the 4 highest redshift objects from our sample. For 3 objects, we did not detect any optical emission in our deep ($t_{int}=1-1.5$~h) Keck spectra, while the remaining 3 objects have a faint optical continuum down to 4800~\AA, but no detectable emission or absorption lines. In Paper~III, we discuss the possible nature of the objects lacking redshift determination. 

Although we obtained redshifts for only 39\% of the $K>20$ objects with small radio sizes, they are significantly more distant than the other objects in our sample. This suggests that an angular size cutoff of $\simlt 5\arcsec$ could also be used as an additional filter to select the highest redshift objects, prior to the $K-$band imaging. Such a cutoff has been used in several samples designed to find HzRG, such as the 6C$^*$ USS sample of \citet{blu98} and the MG sample of \citet{ste99a}. However, although the largest radio galaxy at $z>4$ is 4\farcs3 \citep[8C~1435+635;][]{lac94}, 5 out of the 16 known $3<z<4$ radio galaxies are $>$10\arcsec\ \citep{deb00b}, including 4C~41.17 at $z=3.8$, which has a radio size of 13\arcsec. It is therefore likely that applying such an angular size cutoff will also exclude a number of $z>4$ radio galaxies, and introduce additional selection effects.

\subsection{Radio--near-IR alignment}

Figure \ref{radioKpa} shows a histogram of the difference in orientation between the position angle of the radio structure and the orientation of the near-IR image, as determined from the major axis of the outer fitted isophote in the $K-$band image (\S 3.3). There is a clear trend for the radio and {\it observed} $K-$band morphologies to align. This trend is more obvious when we consider only the radio sources larger than 5\arcsec, whose radio position angle can be measured more accurately (shaded histogram in Fig. \ref{radioKpa}). There are at least 2 additional objects in our sample that show an even closer correlation between the radio and near-IR morphologies: TN~J1049$-$1258 and WN~J1123$+$3141 ($z=3.217$). This can also been seen in Fig.~\ref{Kcutouts}, where the radio position angle is indicated atop $8\arcsec \times 8\arcsec$ cutout images of the 81 detected objects.

Our result is consistent with the high-resolution NICMOS $H-$band observations of 19 radio galaxies with $1.68 < z < 3.13$ \citep{pen01}, which showed radio structures that are often (although with several notable exceptions) aligned with the near-IR morphologies. In contrast, using $K-$band images from B2/6C sample, \citet{eal97} found no alignment effect, and argued that such a radio - near-IR alignment effect only occurs in the most luminous radio sources. Because our sample contains sources of similar or lower radio luminosity than the B2/6C sample, Figure \ref{radioKpa} leads us to question the significance of the distribution based on only 9 objects shown in Figure~6 of \citet{eal97}. We believe our result is more robust because (i) our sample is 5 times larger, (ii) the seeing in our observations is superior (and the pixel scale is 3 times smaller), and (iii) our limiting magnitude is $\sim$2~magnitudes fainter.

Together with the previous NIRC observations of vB98 and the NICMOS observations of \citet{pen01}, our data provide further evidence for the existence of a radio--near-IR alignment effect. Because our sample contains objects over a wide redshift range, the question of the origin of the observed $K-$band emission arises. The spectroscopic redshifts of $\sim$25\% of the objects with $K-$band images from our sample indicate that $\sim$33\% are at $z<2$, $\sim$33\% at $2<z<3$ and $\sim$33\% at $z>3$ (paper~III). For the objects not yet observed spectroscopically, we estimate redshifts $1<z<3$, based on the Hubble $K-z$ diagram (see \S 4). Therefore, the observed $K-$band most likely samples rest-frame $\sim$5500~\AA\ to $\sim11000$~\AA. 
The absence of strongly nucleated morphologies in the $K-$band images suggests there is no strong direct contribution from the AGN to the $K-$band emission. The images also do not show cone-like morphologies, which may be caused by AGN photo-ionization of regions swept up by the passage of the radio jets, as seen in the H$\alpha$ image of B3~0731+438 \citep{mot00}. 
However, for most sources, we cannot exclude a contribution from strong emission lines, which may influence the {\it observed} near-IR emission in the outer region. In several $z \sim 1$ radio galaxies with high-resolution {\it HST} images, the extended emission lines clearly strengthen the alignment effect, but they cannot account for it completely \citep{lon95}.
Only spatially resolved spectroscopy can determine the exact contribution from emission lines. 
Such a contribution can also be excluded if the redshift is in an interval where there are no strong lines in the observed $K$ or $K_S$ band. Unfortunately, we have determed redshifts of only 40\% of the objects observed (see Table~\ref{imjournal}). One example where we can exclude a significant emission line contribution is TN~J1033$-$1339 ($z=2.425$). In objects like these, stellar light seems to be the most likely origin of the $K-$band emission.

\section{The Hubble K-z diagram}

During the last 2 decades, the Hubble $K-z$ diagram has played an important role in the search for and the study of high redshift galaxies \citep[\eg][vB98]{lil84,eal97}. For example, the first radio galaxy discovered at $z>3$ was selected from a radio sample on the basis of a faint $K\sim 18.5$ magnitude \citep{lil88}. In our sample, we combine this faint $K-$band selection with our radio USS criterion.

\subsection{The $K-z$ diagram for radio galaxies}

Figure \ref{Kzradio} presents a $K-z$ diagram that includes the 22 objects with $K-$band data and redshifts, augmented with data of 3 samples drawn from the literature: the 3C sample \citep{lil84,dun93}, the 6C/B2 sample \citep{eal97,raw01}, and the sample of HzRGs observed with NIRC by vB98. We followed the same photometric procedures as described in \citet{eal97} to ensure that all magnitudes are on the same metric system. This involves correcting the magnitudes measured in an 8\arcsec\ diameter aperture to a standard 64~kpc aperture \footnote{\citet{eal97} used a 63.9~kpc aperture, because the 8\arcsec\ aperture used for most of the $K-$band data in the literature supposedly corresponds to this value at $z=1.0$ in their adopted cosmology ($H_0=50~$km~s$^{-1}$~Mpc$^{-1}$, $\Omega_M=1.0$, and $\Omega_{\Lambda}=0$). In fact, 8\arcsec\ corresponds to 68.1~kpc, but we continue to use 64~kpc for the sake of uniformity in the literature.}, assuming the emission within an aperture of radius $r$ is proportional to $r^{0.35}$. This aperture correction brightens the magnitudes by $0 - 0.1$, depending on redshift. As shown in Figure \ref{Kprofiles}, the 8\arcsec\ diameter aperture is too large for most objects, but no smaller-aperture magnitudes have been published for large samples in the literature.

Our new HzRGs obey the same relation as the previous data, but have a larger scatter than the objects observed by vB98, even though the same observational setup was used in both samples.
We now examine the possible reasons for this.

\paragraph{Contribution from emission lines.} 

Using near-IR spectroscopy of HzRGs, \citet{eal93} argued that emission lines may contribute $0-100$\% to the $K-$band magnitude. 
vB98 avoided this problem by observing their sample using specific $K-$band filters which avoid rest-frame wavelengths including strong emission lines. 
Our observations could be more subject to line contamination because we used only the $K-$ and $K_S-$band, mostly without prior knowledge of the spectroscopic redshift. However, the only lines that are expected to contribute significantly to the $K-$band magnitude are \OIIfull, \OIIIfull, and \Ha\ \citep[see][]{mcc93}. This would lead to an asymmetric scatter towards brighter magnitudes around $z\sim$ 4.9, 3.4, and 2.3. For example, \citet{arm98a} found that the \OIII\ lines in 4C~+19.71 at $z=3.594$ contribute 34\% to the $K-$band magnitude. 
However, only one object in our sample shows a clear excess magnitude at these redshifts (WN~J1123$+$3141), but this is an exceptional, complex merging system, in which enhanced star-formation may well contribute more to the $K-$band luminosity in the continuum than optical line emission (\OIIIfull). Moreover, because our sample contains less radio powerful sources, the emission lines are also fainter, due to the correlation of emission line luminosity with radio power \citep{raw91,jar01b}. We therefore dismiss this possibility.

\paragraph{Direct contribution from the AGN.} 

According to unified models, the hosts of radio galaxies harbor hidden AGN which are shielded from our direct view by thick obscuring material, while in quasars there is a direct line of sight toward the unresolved central AGN. Because the AGN emission is extremely bright relative to the emission from the stellar population in the host galaxy, even a small partially obscured contribution could have a significant influence on the total integrated $K-$band magnitude. However, this contribution would appear unresolved and in \S 3.3 we found no evidence for such a component in the $K-$band images. Although our images are not of sufficient quality to put strong limits on a possible direct AGN contribution, we would have clearly detected cases where this component would dominate the $K-$band emission. 

The low selection frequency (325 or 365~MHz) of our USS sample also tends to avoid objects which are dominated by a strong flat spectrum radio core \citep[\eg][]{blu98}. Those objects are thought to be more Doppler boosted due to a decrease in the angle between the jet-axis and the line of sight. Such small viewing angles would increase the chance of a direct line of sight towards the central AGN like in quasars.

At $z\sim 1$, there is also observational evidence from studies of other samples of radio galaxies that the contribution of AGN emission to the $K-$band emission is small. The results of thermal IR observations of 3C and lower redshift radio galaxies by \citet{sim99} and \citet{sim00} show that even in the most powerful radio sources, the contribution by AGN emission to the $K-$band magnitude is $\simlt$10\%. Near-IR polarimetry observations by \citet{ley98} indicate that scattered quasar light makes only a small contribution at near-IR wavelengths, even in sources where this contribution is important in the UV and optical.

\subsection{Relation between radio power and $K-$band magnitude at given redshifts.}

Based on the observation that the $K-$band magnitudes of the stellar populations of $z \sim 1$ 3CR radio galaxies are brighter than those of the lower radio power 6C radio galaxies, \citet{bes98} argued that the stellar masses of the more powerful radio galaxies are greater.  The most powerful radio sources would possess radio beams with kinetic powers close to the Eddington limiting luminosity of a central super-massive black hole. The recently found correlation between black hole mass and galaxy velocity dispersion in nearby elliptical galaxies suggests a causal connection between the formation and evolution of their central black hole and bulge \citep{geb00}. This may also imply that the mass of the black hole scales with the mass of the host galaxy, in which case the most powerful radio sources will contain more massive central engines together with their larger stellar mass. Additional evidence for the existence of such a relation was provided recently by \citet{lac00} who found that the {\it rest-frame} $R-$band magnitudes $M_R$ of $z \sim 1$ radio galaxies increase with radio luminosity. 

To examine the relation between radio power and $K-$band magnitude quantitatively, we first fit a linear relation to the $\log_{10}(z)$ vs.\ $K_{64kpc}$ relation. This line, shown in Figure \ref{Kzradio}, does not have any direct physical meaning, and only serves to measure the deviation from the main trend at a certain redshift.
We determined the radio power at 1.4~GHz rest-frame in a consistent way, using flux densities from the WENSS 325~MHz \citep{ren97}, Texas 365~MHz \citep{dou96}, and NVSS 1.4~GHz \citep{con98}; see \citet{deb00b} for more details. This procedure provides radio powers for 99 radio galaxies with known $K-$band magnitudes.
Figure \ref{Kzpower} plots this deviation from the $K-z$ relation against the radio power at 1.4~GHz. The statistical significance of this correlation is $99.94$\%, with a Spearman rank correlation coefficient of $r=-0.34$. Considering only the 87 sources at $z<2$, the significance drops to 95\%, while for the 12 $z>2$ sources, the significance drops to 88\%. Of course the linear fit to the $\log_{10}(z)$ vs.\ $K_{64kpc}$ relation relation is a serious over-simplification which may well introduce an artificial dependence in Figure~\ref{Kzpower}. However, fitting a second order polynomial to the $K-z$ relation has a minor (positive) influence on the significance of the deviation $K-z/P_{1400}$ relation.
We note that the data used to determine this correlation are either from complete samples of radio surveys, or from our sample, and therefore have complete $K-$band information. So there is no selection effect against faint $K-$band sources. Furthermore, we have concentrated on such fainter $K-$band sources in our spectroscopic campaign (paper~III).

We conclude that at $z\sim 1$, the relation seems to be well established \citep[see also ][]{eal97}, but a larger sample of less radio powerful $z>2$ radio galaxies is needed to confirm the extention to higher redshifts. 
However, if the relation holds, it could provide an explanation for the higher scatter in our sample, compared to the sample observed by vB98 (both observed with NIRC). Our new USS sample is based on radio surveys that are $10-100$ times fainter than the ones used to find the other HzRGs \citep[see Fig. 1 of][]{deb00b}, and therefore includes radio sources with a much larger range of radio luminosities, while the vB98 sample mainly contains radio galaxies with $P_{1400} > 10^{36}$~erg~s$^{-1}$Hz$^{-1}$. 
Because radio power and emission line luminosity in HzRGs are correlated \citep[\eg][]{raw91,mcc93,deb00b}, the emission line luminosities are expected to be weaker in our sample. 
This could also introduce a spectroscopic bias against less powerful radio sources, as it is harder to determine their redshifts. The $K-$radio power relation would then propagate this bias towards brighter $K-$band magnitudes.
Our sample is less subject to this because we have obtained deep spectra with the Keck telescope, while the vB98 sample was based on previous USS samples which used only 3$-$4m class telescopes for optical spectroscopy.

The connection between the radio and emission line luminosities is most likely established through a mutual dependence on the black hole mass. This can also explain the small range in $K-$band magnitudes through the dependence of the host galaxy stellar mass on the same black hole mass \citep{bes98,lac00}.
Assuming the radio power -- host galaxy mass relation is the main determinant of the trends with radio power in the $K-z$ diagram, this would argue that the most massive galaxies at the highest redshifts can be found by selecting the most powerful radio sources.

\subsection{Comparison with radio-quiet galaxies}

\subsubsection{$K-z$ diagram}

For almost two decades, this tight correlation in the Hubble $K-z$ diagram has been used as one of the strongest indications that radio galaxies are identified with massive ellipticals at high redshift. However, this argument is based on the extrapolation from the $z \simlt 1$ observations, where this is indeed the case \citep[\eg][]{lil84}.
A direct comparison between the stellar populations of radio-loud and radio-quiet galaxies at $z \simgt 1$ has not been possible until only very recently. This situation has now changed with the availability of extremely deep imaging and spectroscopy of the Hubble Deep Field \citep[HDF-North;][]{wil96}.

\citet{dic01} obtained a $t_{int}=22.9$h $K-$band image of the HDF-North using the IRIM camera at the Kitt Peak National Observatory 4m telescope in 1996 April. These images reach a formal $5\sigma$ limiting magnitude of $K=21.92$ (2\arcsec\ diameter aperture), which is comparable to or fainter than our NIRC HzRG images. 
M.\ Dickinson kindly provided us with a list of all 185 objects in the HDF-North with known spectroscopic redshifts, and their $K-$band magnitudes or limits.
Because the $K-$band HDF-North image has much poorer resolution ($\sim$ 1\arcsec\ seeing) than the space-based {\it HST} images, the magnitudes of the individual objects were not measured in standard apertures. Instead, only objects detected in the {\it HST} NICMOS F110W ($J-$band) and F160W ($H-$band) images \citep{dic99} were selected, convolved to match the ground-based $K-$band PSF, and then fit to the $K-$band data. In this way overlapping objects are deblended, and meaningful fluxes and limits to every object detected in the NICMOS data can be assigned, regardless of whether it can be detected or not in the ground-based $K-$band image. 
These magnitudes can be considered as approximately total magnitudes, and are calibrated to the same Vega-based photometric system, as also used for the observations of the radio-loud galaxies. The 185 spectroscopic redshifts have been compiled by M.\ Dickinson from identification campaigns of various groups. They are therefore a heterogeneous mix, and do not intend to be complete in any way. The highest redshift galaxy in the sample is HDF 4$-$473.0 at $z=5.60$ \citep{wey98}.

Because of the small survey area, the HDF-North catalog contains few $z \simlt 0.3$ galaxies. To increase the number of radio-quiet galaxies at low redshifts ($z \simlt 1$), we used the redshifts and $K-$band magnitudes of the Hawaii survey \citep{cow94b,son94}. 
These magnitudes were measured in a 3\arcsec or 3\farcs5 diameter circular aperture, and then corrected to total magnitudes using an average correction determined from all bright isolated objects in the field \citep[see][ for details]{cow94b}. 
\citet{son94} provide spectroscopic redshifts for the $K<20$ objects. The highest redshift galaxy in this sample is at $z=1.154$, and the list also contains a $z=2.33$ BALQSO \citep{cow94a}, which we include for completeness, although it can not be directly compared with the other radio-quiet or radio-loud galaxies.

Because the 64~kpc metric apertures determined for the radio galaxies in Figure \ref{Kzradio} are also our best estimate for their total magnitudes, we can now directly compare the $K-$band magnitudes of radio-quiet and radio-loud galaxies over the entire redshift range $0 < z < 5.6$.
Figure \ref{Kz} shows this composite Hubble $K-z$ diagram of radio-quiet and radio-loud galaxies. The most obvious difference with Figure \ref{Kzradio} is the almost complete absence of a relation in the radio-quiet objects. 
There also appears to be a deficiency of $K>20$ objects at $z \simlt 0.3$, but this is due to the limited survey area of the HDF, and the $K<20$ limit of the spectroscopy of the Hawaii survey. This is also the most likely explanation for the absence of bright radio-quiet galaxies. Also note the dearth of sources in the 'redshift desert' at $1.5 \simlt z \simlt 2$, where no bright emission lines are observable with optical spectrographs.

Figure \ref{Kz} shows that at $z \simlt 1$, the radio-loud galaxies trace the bright envelope of the radio-quiet galaxies, while at $z \simgt 1$, they are $\simgt 2$ magnitudes brighter. 
Provided the non-stellar contributions to the $K-$band emission are negligible (as argued above), this is some of the strongest evidence to date that powerful radio galaxies pinpoint the most massive systems out to the highest redshifts currently accessible. Although mass estimates are very uncertain, \citet{bes98} estimate masses exceeding 10$^{11}$M$_{\odot}$ for $z\sim 1$ 3CR radio galaxies. However, this is consisent with other determinations: \citet{dey96} derive a kinematically determined mass estimate of $\sim 8 \times 10^{10}$M$_{\odot}$ for the central region of the $z=0.81$ radio galaxy 3C~265.
These values are an order of magnitude smaller than the ones determined for HDF field galaxies: \citet{pap01} derive masses of $\sim 10^{10}$M$_{\odot}$ for $2.0 \simlt z \simlt 3.5$ Lyman break galaxies in the HDF-North. 
This illustrates that the large difference in survey area between the radio and deep optical surveys can allow the detection of such rare massive objects.

\subsubsection{$R-K /z$ diagram}

The $R-K / z$ diagram has been widely used to argue that the stellar population in radio galaxies is dominated by an evolved population formed at high redshifts ($z_{form} \simgt 5$). At $z \simgt 1$, the scatter increases due to the $k-$correction effect shifting contributions from younger starbursts into the {\it observed} $R-$band \citep[\eg][]{dun89}. Especially at the highest redshifts ($z > 3$), there is evidence for the presence of substantial amounts of star-formation \citep[\eg][]{dey97,pap00}. However, these recent starbursts (possibly induced by the radio activity) cannot produce the majority of the stellar population, as this would produce too large a scatter in the $K-z$ diagram \citep{roc90}. An additional explanation for the increased $R-K$ scatter at $z \simgt 1$ could be the increasing contributions from non-stellar components. This is supported by spectropolarimetry observations suggesting that a scattered quasar component dominates the UV spectra of radio galaxies at $z\sim 1$ \citep[\eg][]{tra98} and $z\sim 2.5$ \citep{ver01}.

To examine if our radio galaxies differ in any way from field galaxies, we also compiled the $R-K$ colors of radio and HDF galaxies. Because our sample contains only 14 spectroscopically confirmed objects with $R-K$ information, we supplemented our sample with 130 galaxies from the MRC 1Jy survey \citep{mcc99} and 41 galaxies from the 7CRS \citep{wil01b}. While all these radio galaxies have spectroscopic redshifts, this does not mean they are an unbiased sample. \citet{wil01a} have made photometric redshift estimates for the 7 sources in their 7CRS sample which did not yield redshifts after optical spectroscopy, and found that 6 of them have $R-K > 5.5$. This is consistent with the results from Keck spectroscopy of our USS sample \citep{deb01}: of the 5 sources that did not yield a redshift after $\sim$1h exposures, one has $R-K=5.8$, while the others have limits $R-K>4$.

For the HDF, we use the photometry and photometric redshifts of \citet{fer99}. To obtain $R-$band colors, we interpolated between the mean wavelengths of the F606 (6031~\AA) and F801 (8011~\AA) filters to 6940~\AA.
We also include the color evolution predicted from stellar evolution synthesis models, calculated with PEGASE~2 \citep{fio97,fio99}. The 5 model grids with $z_{form}=2,3,4,5,6$ represent an instantaneous burst, using a Salpeter IMF ($0.1<{\rm M}<120$M$_{\odot}$, solar metallicity, and no infall.

Figure \ref{zRmK} shows the combined $R-K/z$ diagram of radio and optically selected galaxies.
At $z \simlt 2$, the radio galaxies are clearly redder than the field galaxies, and closely follow the predictions from evolved stellar populations. At higher redshifts, the scatter increases, which can be explained by different competing factors, including: (i) the increasing contribution of young stars and non-stellar emission in the $R-$band, (ii) the incompleteness of the spectroscopic redshift information which tends to be more significant for the reddest objects \citep{wil01a}, and (iii) reddening due to dust \citep{arm98a,arc01,reu01}.
Because the amount of dust emission appears to increase with redshift \citet{arc01}, it is unlikely that the redder colors at $z \simlt 2$ are due to reddening by dust. We therefore conclude that, $z \simlt 2$ radio galaxies appear clearly redder than the HDF field galaxies, and interpret this as evidence for the more evolved stage of their stellar populations. At $z \simgt 2$, a detailed decomposition of the different components contributing to the $R-$ and $K-$band emission would be needed to make a proper comparison between both classes.

Some of the field galaxies have colors similar to those of the radio galaxies, indicating some field galaxies also contain evolved stellar populations, or are very dusty \citep[\eg][]{arm98b,cim98,dey99b}. The fact that almost no HDF galaxies have colors significantly redder than the radio galaxies supports the idea that the latter formed at higher redshift. However, the larger scatter in $R-K$ color for the HDF galaxies does not imply that they are not ellipticals, as morphologically selected samples from the HDF have found bluer field ellipticals \citep[\eg][]{sch99,dic00}.

\subsubsection{The fraction of massive galaxies which harbour powerful radio sources}

What fraction of massive galaxies at high redshift do powerful radio sources represent? To answer this question, a large-area spectroscopic survey of $K\sim 20$ galaxies, with follow-up radio observations would be needed. Although the Hawai'i survey covers 250 arcmin$^2$ for $K<16$ objects, this area is limited to 5 arcmin$^2$ for $K<20$ \citep{son94}, similar to the coverage of the HDF. This is too small to detect statistically significant numbers of galaxies at the highest redshifts. 

Nevertheless, it is worth pointing out that the $z>4$ galaxy in the HDF with the brightest radio emission ($S_{1400}=470 \mu$Jy) is more than a magnitude brighter in $K-$band than the other 6 at $z>4$ \citep[VLA~J123642+621331 at $z=4.424$, ][]{wad99}. There is now circumstantial evidence that this object contains an AGN, including (i) the detection of an unresolved component in a 0\farcs026 resolution radio image obtained with the European VLBI Network (EVN) \citep{gar01}, and (ii) the detection of X-ray emission with {\it Chandra} which is consistent with the optical to X-ray flux ratio seen in local AGN \citep{bra01}. 

At lower redshift, VLA~J123644+621133 is a $z=1.050$ FR~I source \citep{ric98}. The $K=16.92$ of this object is comparable to the the brightest 3C sources at these redshifts. 
Several other AGN in the HDF detected in deep radio and/or {\it Chandra} X-ray maps \citep{hor00} inhabit fairly normal galaxies. One of the {\it Chandra} sources with a photometric $z_{phot}\simeq 2.6$ is an extremely red object with $K=22.07$, and is likely an obscured AGN (M.\ Dickinson, private communication). Such obscured AGN at $z \simlt 2$ might also be present in our radio selected sample (see paper~III).

To summarize, radio galaxies remain the most efficient tracers of the most massive galaxies at the highest redshifts accessible with optical spectrographs. Some of the most luminous objects at a given redshift in the optically selected surveys also appear to harbour an AGN, but the present sample is too small to determine the fraction of massive galaxies that contains an AGN. 
Moreover, deep {\it Chandra} X-ray observations indicate that a substantial fraction of AGNs might be heavily obscured \citep{mus00}, and might prove difficult to identify at optical or near-IR wavelengths.
Because the life-time of a powerful radio source \citep[$10^7 - 10^8$~y;][]{blu99b} is short compared to the Hubble time, a substantial fraction of massive galaxies should be going through a period of radio inactivity, and can only be found from non-radio selected surveys.
An approach made possible with the advent of deep X-ray surveys with {\it Chandra} and {\it XMM-Newton} is to use the X-ray emission as a signpost of nuclear activity. This provides a powerful alternative for radio observations, although the survey areas covered are still small.
From deep multi-wavelength observations of the Hawai'i Deep Survey Field SSA13, \cite{bar01} find that $7^{+5}_{-3}$\% of the optically luminous galaxies ($-22.5 > M_I > -24$) are indeed X-ray sources.

\section{Repercussions for HzRG searches}

The original goal of the imaging programme was to identify our USS sources as promising high redshift radio galaxy targets for subsequent optical spectroscopy. We find that due to the red $R-K$ colors of the host galaxies, near-IR imaging is a far more efficient technique for obtaining HzRG identifications than optical imaging. Near-IR imaging leads to an almost complete identification rate, and provides an approximate photometric redshift by means of the Hubble $K-z$ diagram, or, potentially, from a modified $K-z-P_{\rm radio}$ relation. Figure \ref{Khis} shows that the majority of the $K-$band images are relatively bright, suggesting that the average redshift of our sample is relatively moderate. To examine this quantitatively, we first transformed the 8\arcsec\ diameter apertures to 64~kpc metric apertures using the average correction of $K_{64kpc} = K(8\arcsec) + 0.2$. We then used the fit of the $K_{64kpc}$ versus $\log_{10}(z)$ relation as obtained in \S 4 to determine the predicted redshift of these objects. Although this is a severe over-simplification which ignores the scatter in the $K-z$ relation, and is based on only 13\% of the total number of sources in our USS sample, it can provide a first estimate of the redshift distribution of the 81 objects detected at $K-$band.

Figure \ref{zpredhis} shows this predicted redshift distribution. The median predicted redshift from this sample is 2.06, while the mean is $\bar{z}=2.25$. This median redshift is consistent with the rather steep surface brightness profiles typical of $z<3$ radio galaxies (see \S 3.3).
Figure \ref{zpredhis} also shows the distribution of the spectroscopic redshifts obtained, which has a mean/median at $z \simeq 2.5$. This difference illustrates the advantage of using an additional faint near-IR magnitude selection criterion to increase the chances for finding very high redshift radio galaxies. 

The distributions in Figures \ref{R-Khis} and \ref{zpredhis} suggest that the majority of the sources in our sample are very red objects or even EROs at $0.5 \simlt z \simlt 2.5$ with a tail of very high redshift radio galaxies. Wide-field surveys of EROs have indeed shown that the density of ERO increases with $K-$band magnitude out the their completeness limit of $K\sim 19$ \citep[\eg][]{dad00}. Because the imaging sub-sample selects objects at $K-$band, and a number of objects were even pre-selected to have faint $R-$band magnitudes, our sample includes a large number of very red objects and EROs that happen to be radio-loud. This population might represent more than half of our total sample, suggesting that very red objects at $z\sim 1.5$ are probably important contributors to galaxy counts in steep-spectrum radio-selected samples. These red objects generally have $18<K<20$, explaining our increased success in finding $z>3$ radio galaxies by concentrating on the $K>20$ identifications.

\section{Conclusions}

We have obtained 83 optical images, and 85 near-IR images of sources selected from our USS sample. Our main conclusions from these observations are:

$\bullet$ $\sim$50\% of the objects are detected in the optical images down to $R\sim 24-25$.

$\bullet$ 94\% of the objects are detected in the near-IR $K-$band down to $K\sim 22$, with a mean $\bar{K}=19.26$ in a 2\arcsec\ diameter aperture. One of the non-detections is due to the limited sensitivity of the radio map, while the other 3 are intrinsically faint $K-$band sources, 1 having $K\gtrsim 19$, and 2 having $K\gtrsim 22$.

$\bullet$ The distribution of $R-K$ colors shows that at least 1/3 have $R-K>5$, and at least 4 objects ($>$10\%) are EROs with $R-K>6$.

$\bullet$ The $K>20$ identifications appear only associated with small ($\simlt 2$\arcsec) radio sources.

$\bullet$ The $K-$band morphologies show mostly compact objects typical of $z \simlt 3$ radio galaxies.

$\bullet$ The major axes of the $K-$band morphologies, as determined from the fitting of ellipsoidal isophotes are preferentially oriented along the radio axes. This provides further evidence for the existence of a near-IR/radio alignment effect.

$\bullet$ The 22 objects from our sample with known $z$ and $K$ obey the same relationship in the Hubble $K-z$ diagram as previous radio samples, but with a larger scatter.

$\bullet$ The scatter may well be dominated by the radio power, which is correlated with the deviation from a linear fit to the $\log_{10}(z)$ vs.\ $K_{64kpc}$ relation with a 99.94\% significance level. This indicates that the most powerful radio sources are located in the most massive host galaxies, probably due to a mutual correlation if galaxy mass and radio power on the mass of the central black hole. The previously known HzRGs were drawn from samples which included only the most powerful radio sources, giving rise to a tighter correlation in the $K-z$ diagram.

$\bullet$ A comparison between the $K-z$ relation of radio-loud and radio-quiet galaxies determined from the HDF-North and the Hawaii survey shows that the radio-loud galaxies define the luminous envelope of the $K-$band magnitudes at $z \simlt 1$, while at $z \simgt 1$, the radio-loud galaxies are $\simgt 2$ magnitudes brighter. This is amongst the strongest evidence to date that radio galaxies trace the most massive forming stellar populations at high redshifts.

$\bullet$ Radio galaxies generally have redder $R-K$ colors than optically selected galaxies from the HDF, especially at $z \simlt 2$. This indicates radio host galaxies also contain more evolved stellar populations.

$\bullet$ Some of the brightest objects in the K$-$band from the optically selected surveys appear to contain AGN, suggesting that the fraction of massive galaxies containing AGN is quite high. However, a much larger $K-$band selected sample of high redshift galaxies would be needed to estimate this fraction.

$\bullet$ The HzRGs with the faintest $K-$band identifications may be extremely dusty and/or at extremely high ($z > 7$) redshifts. We have therefore embarked on a program to obtain near-IR spectroscopy at Keck and sub-mm observations at JCMT/IRAM to investigate this in more detail and will report on the results in a forthcoming paper \citep{reu01}.

\acknowledgments

We thank Mike Brown for letting us use a severely crippled NIRC camera
on Keck~I for half a night (1999 September 22) when the instrument was
of no use to his demanding Solar System observations, while it could
be employed for our simple purpose of $K-$band galaxy identifications
in the early Universe.  We thank Mark Dickinson for providing the
catalog of redshifts and $K-$band magnitudes in the HDF-North, and for
helpful comments on the $K-z$ diagram. We thank Roy Gal for obtaining
Palomar 60\arcsec\ images needed for the astrometry of a few of the
fields.  We are grateful for the excellent help provided by the staff
of the Lick, Keck, ESO, CTIO, and WHT telescopes. We thank Damien Le
Borgne, Brigitte Rocca-Volmerange, Michel Fioc and Philip Best for
useful discussions, and the referee Arjun Dey for constructive
comments that have improved the paper.  This research made use of the
NASA/IPAC Extragalactic Database (NED) which is operated by the Jet
Propulsion Laboratory, California Institute of Technology, under
contract with the National Aeronautics and Space Administration. The
work by C.D.B., W.v.B., and S.A.S.\ at IGPP/LLNL was performed under
the auspices of the US Department of Energy by University of
California Lawrence Livermore National Laboratory under contract
W-7405-ENG-48. This work was supported in part by the Formation and
Evolution of Galaxies network set up by the European Commission under
contract ERB FMRX-- CT96--086 of its TMR programme, and by a Marie
Curie Fellowship of the European Community programme ``Improving Human
Research Potential and the Socio-Economic Knowledge Base'' under
contract number HPMF-CT-2000-00721.

\newpage

\begin{deluxetable}{rrccccccc}
\tablewidth{0pc}
\tabletypesize{\small}
\tablecaption{Recent Radio Redshift Surveys}
\tablehead{
\colhead{Survey} & \colhead{Selection in $S$} & \colhead{Selection in $\alpha$} & \colhead{Median} & \colhead{Mean} & \colhead{\% $z_{spec}$} & \colhead{Mean $z$} & \colhead{\# sources} & \colhead{Reference}\\
\colhead{} & \colhead{[mJy]} & \colhead{} & \colhead{$S_{1400}$ [mJy]} & \colhead{$S_{1400}$ [mJy]} & \colhead{} & \colhead{} & \colhead{} & \colhead{}}

\startdata
MRC    & $S_{408} > 950$                  &  All                           &  463 &  641 &  67.0 & 0.74 &  558 & 1 \\
6CE    & 2000$ < S_{151} < $3930          &  All                           &  391 &  427 &  94.9 & 1.14 &\phn59& 2 \\
MP     & $S_{408} > 700$; $S_{4850} > 35$ &  $\alpha_{408}^{4850} < -1.20$ &  311 &  395 &  15.5 & 0.88 &\phn58& 3 \\
USS    & varies\tablenotemark{a}          & $\alpha<-1.0$\tablenotemark{a} &  281 &  318 &\phn2.3& 1.27 & 1165 & 4 \\
7C-III & $S_{151} > 500$                  &  All                           &  184 &  211 &  81.5 & 1.19 &\phn54& 5 \\ 
6C$^*$ &  960$ < S_{151} < $2000          &  $\alpha_{151}^{4850} < -0.98$ &  131 &  127 & 100.0 & 1.96 &\phn29& 6 \\
TN     & $S_{365} > 150$; $S_{1400} > 10$ &  $\alpha_{365}^{1400} < -1.30$ &\phn74&  118 &\phn7.1& 2.10 &  268 & 3 \\
WN     & $S_{325} >  18$; $S_{1400} > 10$ &  $\alpha_{325}^{1400} < -1.30$ &\phn18&\phn41&\phn4.4& 1.87 &  343 & 3 \\
\enddata
\tablenotetext{a}{The USS sample of R\"ottgering \etal (1994)\nocite{rot94} consists of 9 sub-samples, each with different flux density and spectral index limits. See Table 4 of R\"ottgering \etal (1994) for details.}
\tablerefs{(1) \citet{mcc96}; (2) \citet{raw01}; (3) \citet{deb00a}; (4) \citet{rot94}; (5) \citet{lac99}; (6) \citet{jar01a}}  
\label{surveys}
\end{deluxetable}

\begin{deluxetable}{llrrcr}
\tablewidth{0pc}
\tabletypesize{\scriptsize}
\tablecaption{Observing Sessions}
\tablehead{
\colhead{Run} & \colhead{UT Date} & \colhead{Telescope} & \colhead{Instrument} & \colhead{Photometric?} & \colhead{Seeing}}

\startdata
L3 & 1996 Apr 19     & Lick 3m  & Kast           & no     & 1\farcs8 \\
L6 & 1996 Nov 11     & Lick 3m  & Kast           & yes    & 1\farcs3 \\ 
L7 & 1997 Jan 7-9    & Lick 3m  & Kast           & no     & 2\arcsec-3\arcsec \\
E2 & 1997 Mar 5-6    & ESO 3.6m & EFOSC1         & yes    & 1\farcs3 \\
E3 & 1997 Apr 9-11   & ESO 3.6m & EFOSC1         & yes    & 1\farcs4 \\
W4 & 1997 May 1-2    & WHT 4.2m & Prime-focus    & yes    & 0\farcs9 \\
L8 & 1997 May 6-7    & Lick 3m  & Kast           & yes    & 1\farcs7 \\
L9 & 1997 Jun 6-8    & Lick 3m  & Kast           & yes    & 1\farcs1--2\arcsec \\
L10 & 1997 Aug 8     & Lick 3m  & Kast           & no     & 1\farcs2--1\farcs5 \\
K4 & 1997 Sep 10-11  & Keck 10m & NIRC           & yes    & 0\farcs4--0\farcs5 \\
C1 & 1998 Feb 16     & CTIO 4m  & CIRIM          & yes    & 0\farcs5 \\
K7 & 1998 Apr 18-19  & Keck 10m & NIRC           & yes    & 0\farcs4--0\farcs6 \\
L13 & 1998 Aug 23-24 & Lick 3m  & Kast           & yes    & 1\farcs3 \\
C2 & 1998 Sep 1-4    & CTIO 4m  & CIRIM          & partly & 0\farcs6--1\arcsec \\
L15 & 1998 Oct 18-19 & Lick 3m  & Kast           & partly & 1\farcs0--1\farcs3 \\
K9 & 1998 Dec 19-20  & Keck 10m & LRIS           & yes    & 0\farcs7 \\
W5 & 1999 May 17-19  & WHT 4.2m & Prime-focus    & yes    & 1\farcs0 \\
K10 & 1999 May 22-23 & Keck 10m & NIRC           & yes    & 0\farcs3--0\farcs4 \\
K12 & 1999 Sep 23-24 & Keck 10m & NIRC           & no\tablenotemark{a} & 0\farcs4--0\farcs7 \\
P1  & 1999 Oct 13    & Palomar 60\arcsec & CCD13 & yes    & 0\farcs9 \\
K13 & 2000 Jan 30    & Keck 10m & NIRC           & yes    & 0\farcs6 \\
\enddata
\tablenotetext{a}{Affected by instrumental problems, see \S2.1.1}
\label{imobservations}
\end{deluxetable}

\begin{deluxetable}{rlllr}
\tablewidth{0pc}
\tabletypesize{\scriptsize}
\tablecaption{Journal of Imaging Observations}
\tablehead{
\colhead{Source} & \colhead{$z$} & \colhead{Run} & \colhead{Filter} & \colhead{$t_{exp}$(s)}
}
\startdata
WN J0000$+$4654 &\nodata& K12 & $K         $ &  780 \\
WN J0007$+$3641 &\nodata& L13 & $r_S       $ &  600 \\
WN J0034$+$4142 &\nodata& L13 & $r_S       $ & 3000 \\
                &       & K12 & $K         $ & 1020 \\
WN J0040$+$3857 & 2.605 & L15 & $r_S       $ & 2700 \\
MP J0100$-$6403 &\nodata& C2  & $K         $ & 1800 \\
TN J0102$-$1055 &\nodata& K12 & $K         $ &  780 \\
MP J0114$-$3302 &\nodata& C2  & $K         $ & 1920 \\
\enddata
\label{imjournal}
\end{deluxetable}

\begin{deluxetable}{llccr}
\tablewidth{0pc}
\tablecaption{Observational Setup}
\tablehead{
\colhead{Telescope} & \colhead{Instrument} & \colhead{Filter} & \colhead{FOV\tablenotemark{a}}  & \colhead{Pixel scale}}

\startdata
Keck     & LRIS       & $I$                  & 6\arcmin$\times$7\farcm8     & 0\farcs212\phn $\pm$ 0\farcs001\phn \\
\\                              
WHT      & PFIP+EEV42 & $R$                  & 8\arcmin$\times$16\arcmin    & 0\farcs236\phn $\pm$ 0\farcs001\phn \\
         & PFIP+LOR2  & $R$                  & 9\arcmin$\times$9\arcmin     & 0\farcs2631 $\pm$ 0\farcs0006 \\
\\                              
ESO 3.6m & EFOSC1     & $R$                  & 5\farcm2$\times$5\farcm2     & 0\farcs6075 $\pm$ 0\farcs0007 \\
\\                              
Lick 3m  & Kast       & $r_S$                & 145\arcsec$\times$145\arcsec & 0\farcs789\phn $\pm$ 0\farcs0035 \\
\\                              
Palomar 60\arcsec & CCD13 & $r$              & 12\farcm9$\times$12\farcm9 & 0\farcs3786\phn $\pm$ 0\farcs0005 \\
\\                              
\hline
\\                              
Keck     & NIRC       & $K, K_S, K^{\prime}$ & 38\farcs4$\times$38\farcs4   & 0\farcs1516 $\pm$ 0\farcs0016 \\
\\                              
CTIO 4m  & CIRIM      & $K, K_S$             & 54\arcsec$\times$54\arcsec   & 0\farcs219\phn $\pm$ 0\farcs007\phn \\
\enddata
\tablenotetext{a}{Field of view of a single exposure; the combined images might be significantly larger due to the dithering applied.}
\label{imsetup}
\end{deluxetable}

\begin{deluxetable}{lcll}
\tablewidth{0pc}
\tabletypesize{\footnotesize}
\tablecaption{Astrometric Data}
\tablehead{
\colhead{Source} & \colhead{} & \colhead{$\alpha_{J2000}$} & \colhead{$\delta_{J2000}$}\\
\colhead{} & \colhead{} & \colhead{h m s} & \colhead{\arcdeg \arcmin \arcsec} }

\startdata
WN J0000$+$4654 & Radio     & 00 00 28.88 & $+$46 54 40.7 \\
                & IR        & 00 00 28.93 & $+$46 54 40.9 \\
                & A$^{IR}$  & 00 00 29.25 & $+$46 54 53.8 \\
WN J0007$+$3641 & Radio     & 00 07 02.92 & $+$36 41 55.9 \\
                & Opt       & 00 07 02.80 & $+$36 41 55.8 \\
                & A$^{opt}$ & 00 07 04.92 & $+$36 42 09.3 \\
WN J0034$+$4142 & Radio     & 00 34 29.21 & $+$41 42 09.5 \\
                & IR        & 00 34 29.10 & $+$41 42 15.7 \\
\enddata
\tablecomments{'Radio' indicates radio source position, 'Opt' optical identification, 'IR' near-IR identification, 'A$^{opt}$' offset star within optical image, and 'A$^{IR}$' offset star within near-IR image.}
\label{astrometry}
\end{deluxetable}

\begin{deluxetable}{lrccr}
\tablewidth{0pc}
\tabletypesize{\scriptsize}
\tablecaption{Optical Photometry of USS Sources}
\tablehead{
\colhead{Source} & \colhead{$z$} & \colhead{Run} & \colhead{Filter} & \colhead{Mag}\\
\colhead{} & \colhead{} & \colhead{} & \colhead{} & \colhead{4\arcsec}}

\startdata
WN J0007$+$3641 &\nodata& L13 & $r_S$ &    21.6                   \\
WN J0034$+$4142 &\nodata& L13 & $r_S$ & $>$24.5                   \\
WN J0040$+$3857 & 2.606 & L15 & $r_S$ & 23.3:\tablenotemark{\dag} \\
TN J0121$+$1320 & 3.516 & L7  & $r_S$ & $>$24                     \\
WN J0137$+$3250 &\nodata& L15 & $r_S$ & 23.2\tablenotemark{\dag}  \\
WN J0155$+$8036 &\nodata& L15 & $r_S$ & 22.3\tablenotemark{\dag}  \\
TN J0205$+$2242 & 3.506 & L7  & $r_S$ & $>$24                     \\
TN J0218$+$0844 &\nodata& L10 & $r_S$ & $>$22.5                   \\
\enddata
\tablenotetext{\dag}{Estimated from non-photometric conditions.}
\tablenotetext{a}{Nearby bright star prohibits measurement.}
\tablecomments{Uncertain values are indicated with a colon.}
\label{rphotometry}
\end{deluxetable}

\begin{deluxetable}{rrllrcccccrc}
\newcommand\nph{\tablenotemark{\dag}}
\tablewidth{0pc}
\tabletypesize{\tiny}
\tablecaption{$K-$band Photometry and Morphology}
\tablehead{
\colhead{Source} & \colhead{z} & \colhead{Run} & \colhead{Filter} & \colhead{A(K)\tablenotemark{a}} & \colhead{Mag} & \colhead{Mag} & \colhead{Mag} &  \colhead{$r_{hl}$} & \colhead{Ellipticity} & \colhead{PA} & \colhead{Morphology} \\
\colhead{} & \colhead{} & \colhead{} & \colhead{} & \colhead{} & \colhead{2\arcsec} & \colhead{4\arcsec} & \colhead{8\arcsec} & \colhead{} & \colhead{} & \colhead{} & \colhead{}}

\startdata
WN J0000$+$4654 &\nodata& K12 & $K         $ & 0.041 & 18.49$\pm$0.04     & 17.96$\pm$0.04     & 16.65$\pm$0.04     & 0\farcs3    & \nodata&  147\arcdeg & compact       \\
WN J0034$+$4142 &\nodata& K12 & $K         $ & 0.025 & 18.80$\pm$0.05     & 18.56$\pm$0.05     & 18.52$\pm$0.10     & 0\farcs4    &  0.10  &   36\arcdeg & compact       \\
MP J0100$-$6403 &\nodata& C2  & $K         $ & 0.008 & 19.23$\pm$0.15     & 18.95$\pm$0.17     & 18.61$\pm$0.28     & \nodata     & \nodata& \nodata     & low S/N       \\
TN J0102$-$1055 &\nodata& K12 & $K         $ & 0.010 & 19.45$\pm$0.06     & 19.03$\pm$0.07     & 18.86$\pm$0.12     & 0\farcs8    &  0.48  &  149\arcdeg & compact       \\
MP J0114$-$3302 &\nodata& C2  & $K         $ & 0.009 & 18.59$\pm$0.11\nph & 17.94$\pm$0.10\nph & 17.66$\pm$0.15\nph & \nodata     & \nodata& \nodata     & low S/N       \\
WN J0117$+$3715 &\nodata& K12 & $K         $ & 0.020 & 18.22$\pm$0.04     & 17.57$\pm$0.04     & 17.28$\pm$0.05     & \nodata     & \nodata& \nodata     & companion     \\
TN J0121$+$1320 & 3.516 & K4  & $K_S       $ & 0.014 & 19.47$\pm$0.06     & 19.00$\pm$0.06     & 18.76$\pm$0.07     & \nodata     & \nodata&   85\arcdeg & companions    \\
MP J0130$-$8352 &\nodata& C2  & $K         $ & 0.047 & 18.11$\pm$0.09\nph & 17.97$\pm$0.10\nph & 18.52$\pm$0.31\nph & \nodata     & \nodata& \nodata     & compact       \\
\enddata
\label{kphotometry}
\tablenotetext{a}{Extinction in $K-$band, from \HI\ column densities.}
\tablenotetext{\dag}{Estimated from non-photometric conditions.}
\end{deluxetable}

\newpage

\psfig{file=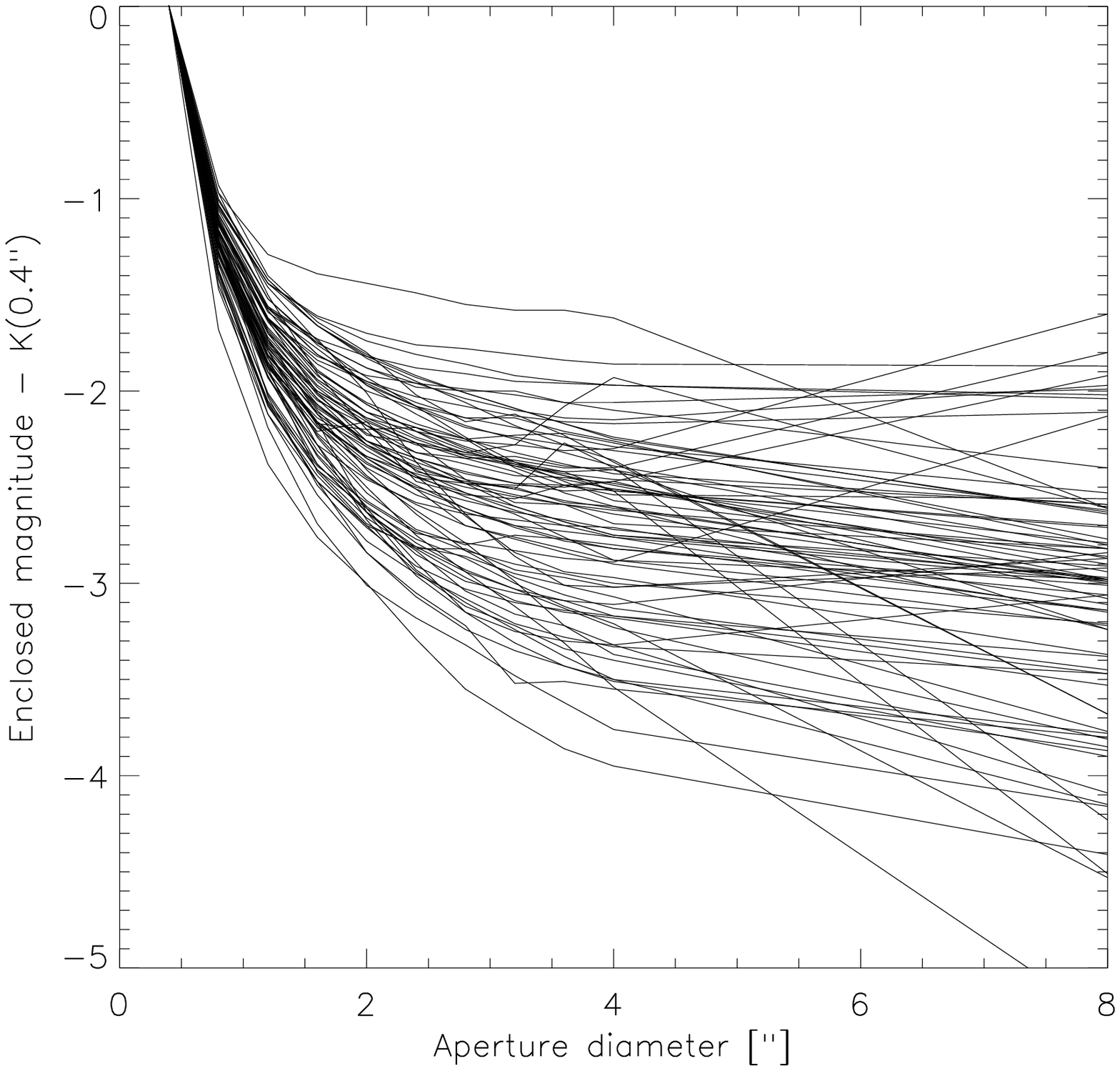,width=16cm}
\figcaption[Kprofiles]{Enclosed $K$-band magnitude as a function
of aperture diameter for our sample of USS radio sources. For the large majority of sources, most of the flux is enclosed in within a 2\arcsec\ aperture. \label{Kprofiles}} 

\newpage

{\bf The individual figures are available on http://www.strw.leidenuniv.nl/$\sim$debreuck/papers/}
\figcaption[Kcutouts]{Cutout images of the 81 $K-$band detections. Each image is $8\arcsec \times 8\arcsec$, and the origin is the position listed with 'IR' in Table \ref{astrometry}. The vector next to the name indicates the position angle of the radio source, unless no high-resolution radio map is known. The images are oriented North up and East left. \label{Kcutouts}}

\newpage
\psfig{file=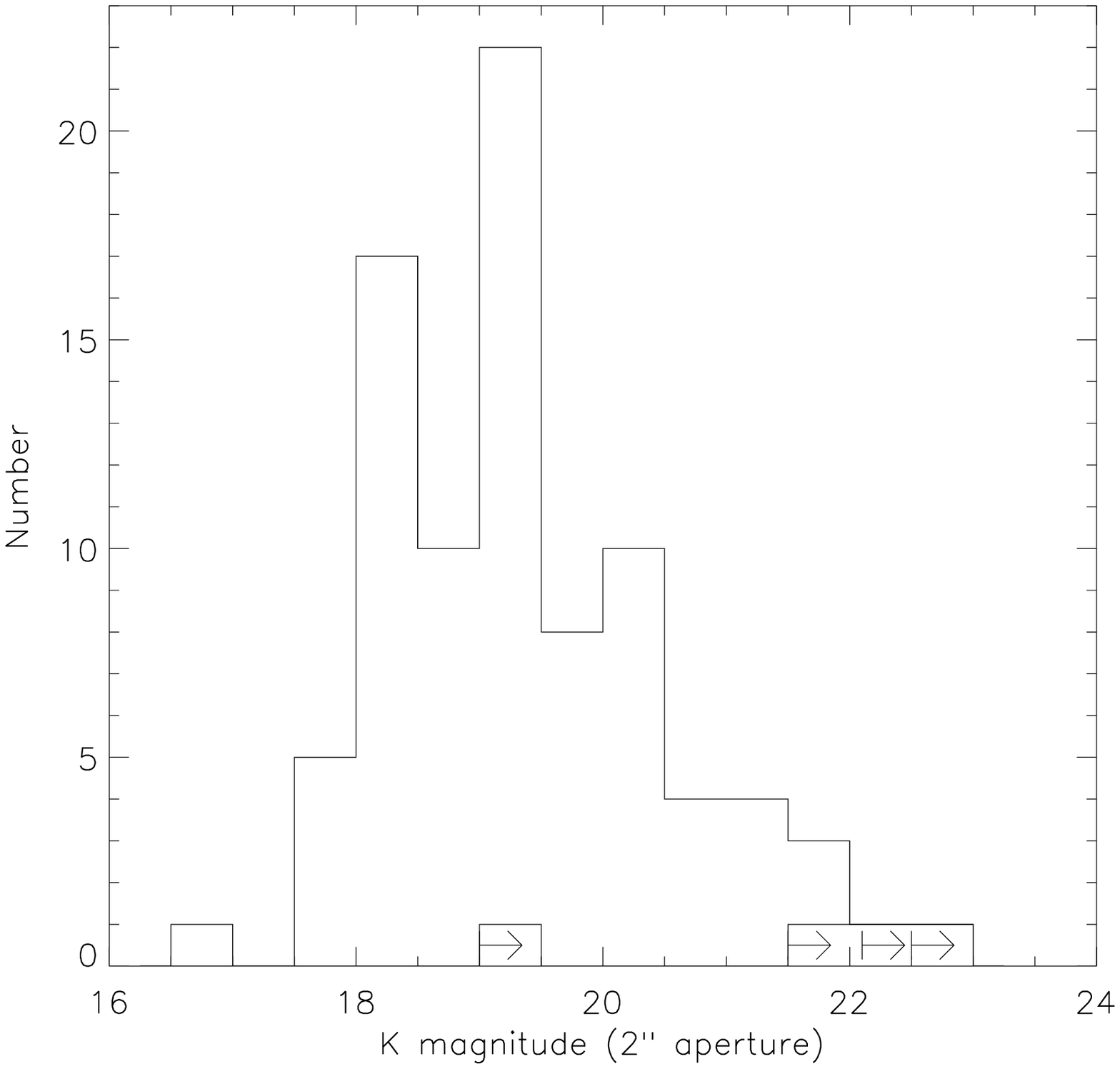,width=16cm}
\figcaption[Khis]{Histogram of the $K-$band magnitudes measured in a 2\arcsec\ diameter aperture. Non detections are indicated by arrows. \label{Khis}} 

\psfig{file=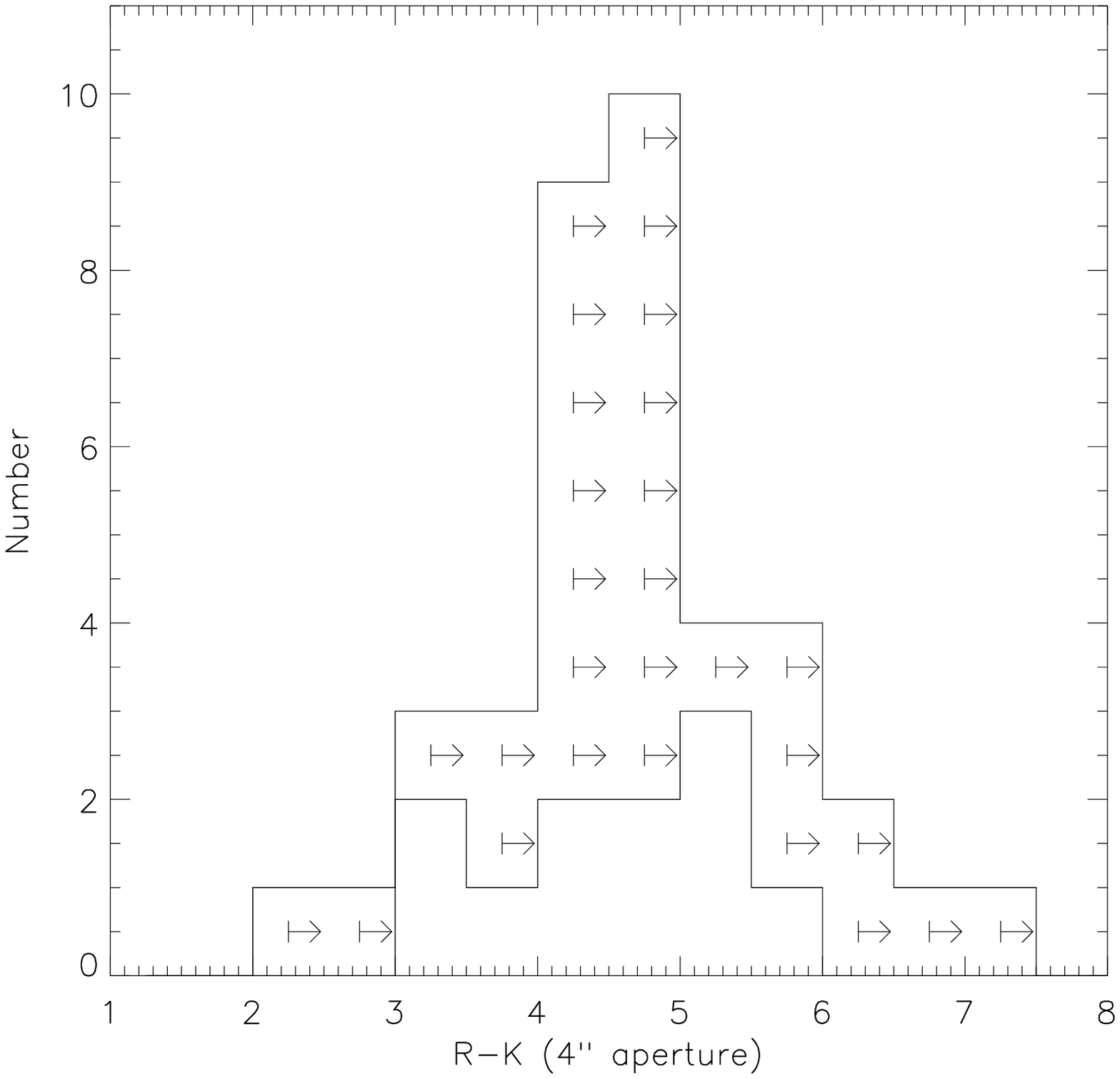,width=16cm}
\figcaption[Khis]{Histogram of the $R-K$ colors measured in a 4\arcsec\ diameter aperture. Non detections in the $R-$band are indicated by arrows. \label{R-Khis}} 

\psfig{file=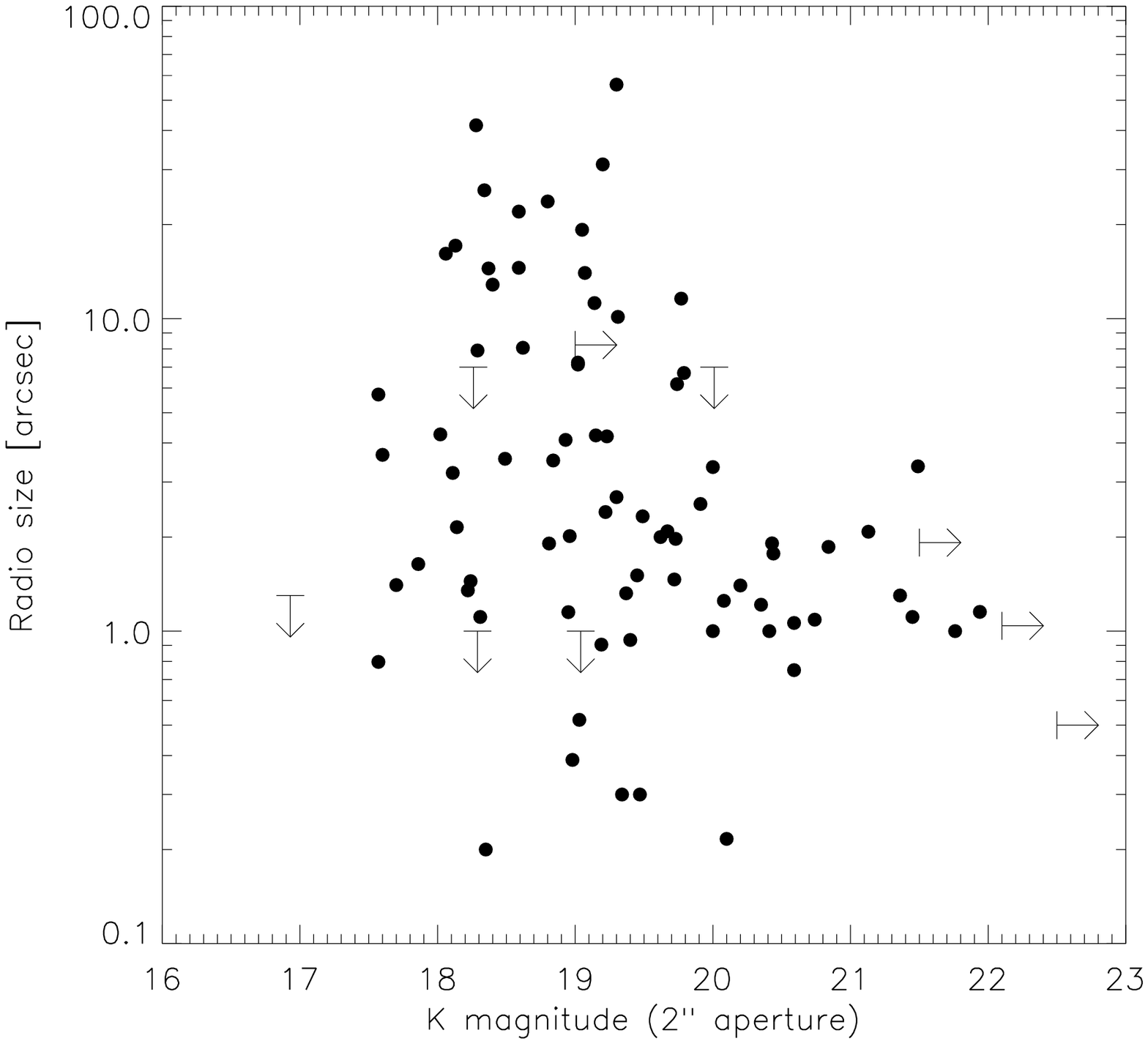,width=16cm}
\figcaption[Klas]{Radio largest angular size plotted against $K-$magnitude in a 2\arcsec\ diameter aperture. Note that all the $K>20$ sources are $\simlt 3\arcsec$, while brighter sources have a larger range of sizes. \label{Klas}}

\psfig{file=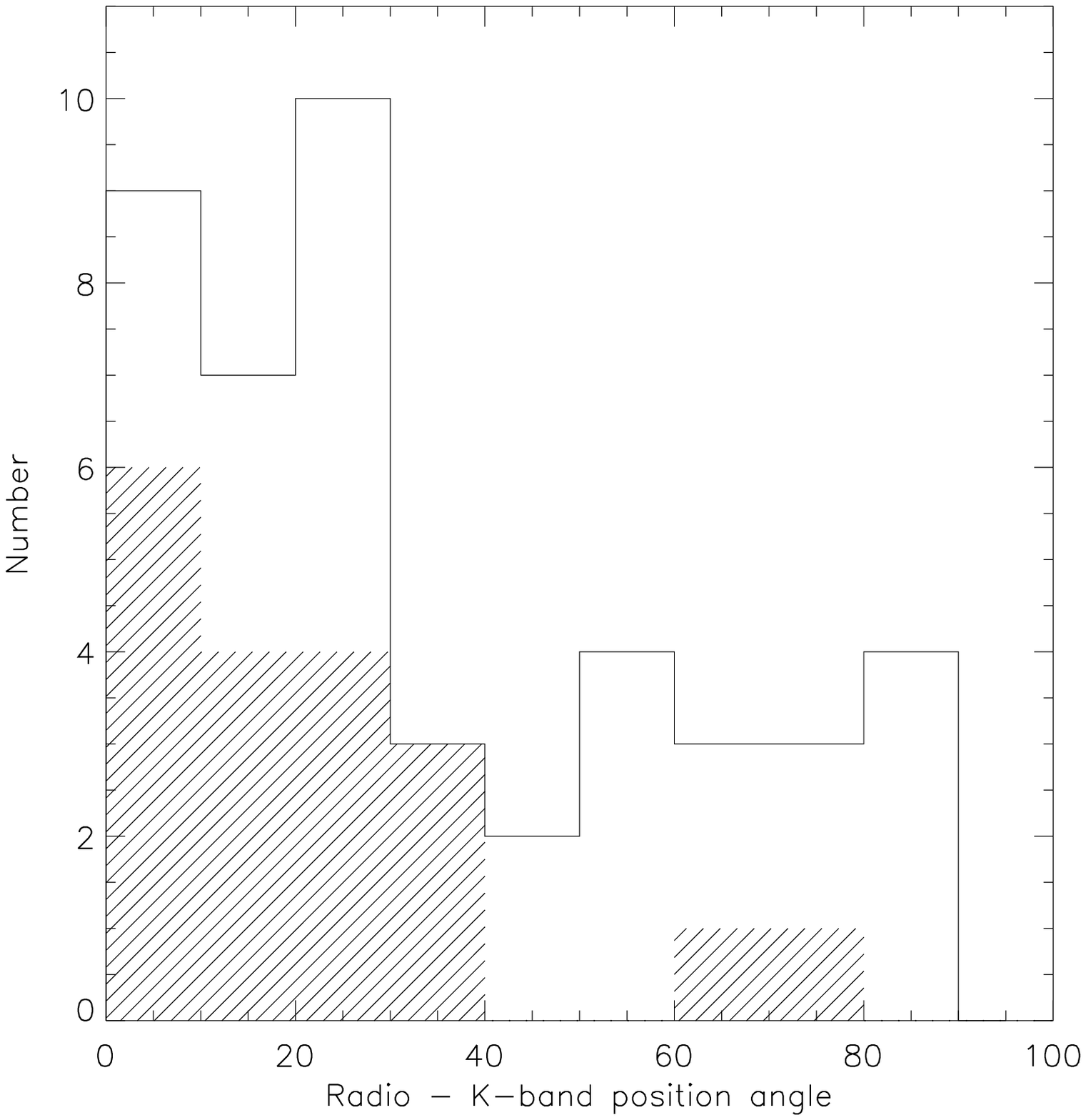,width=16cm}
\figcaption[radioKpa]{Distribution of the differences in position angle between the radio emission and the $K-$band emission of our USS sources, measured at the outer isophote of the {\tt ellipse} fitting. The shaded histogram gives the distribution of the sources with radio sizes larger than 5\arcsec. \label{radioKpa}}

\psfig{file=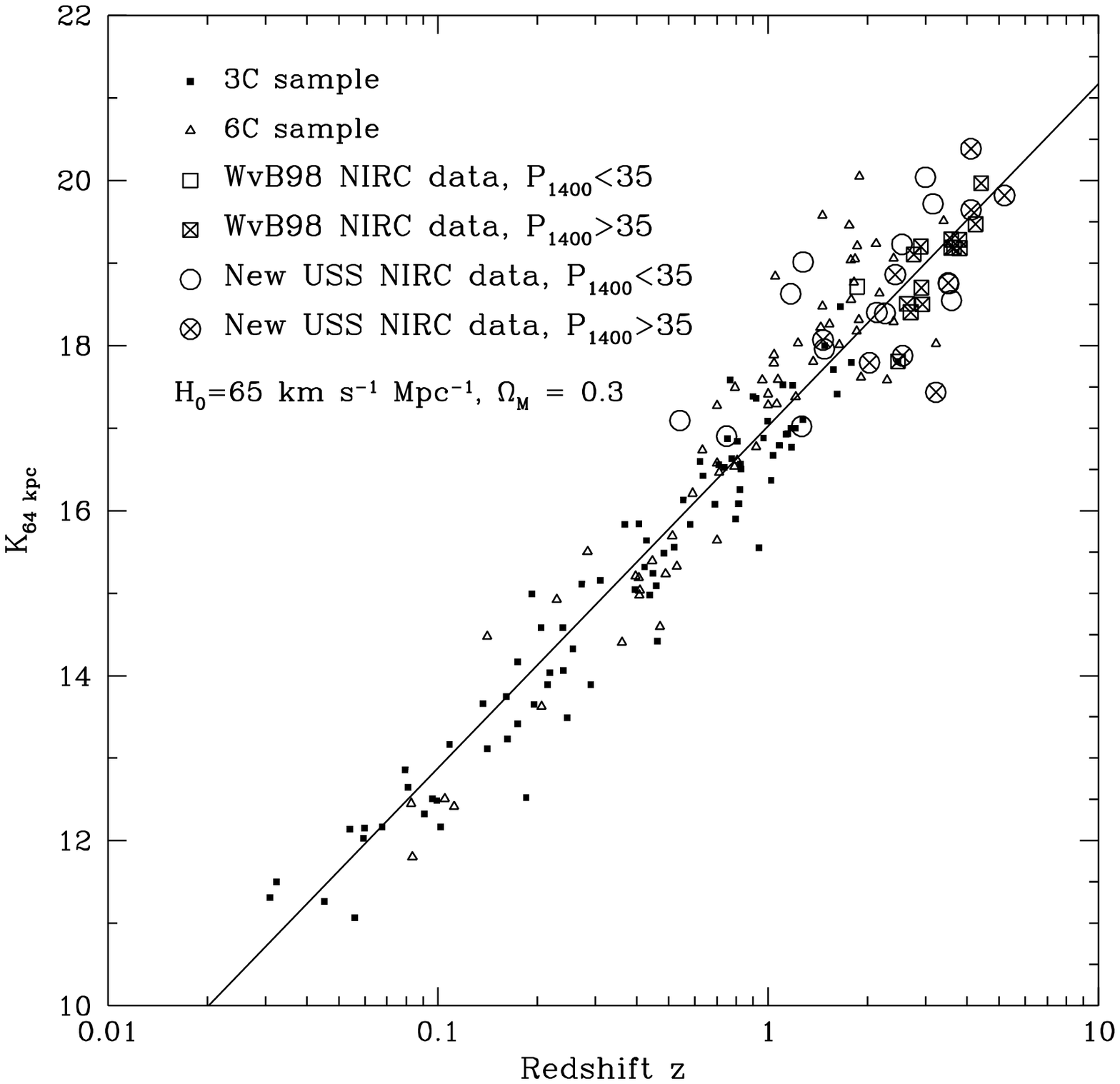,width=16cm}
\figcaption[Kzradio]{Hubble $K-z$ diagram of radio galaxies. Our new USS NIRC data are plotted in the same aperture corrected 64~kpc metric diameter magnitudes as the literature data from \citet{wvb98} and \citet{eal97}. The NIRC data of the sources with radio power $P_{1400}>10^{35}$~erg~s$^{-1}$~Hz$^{-1}$ are filled with a cross. The line shows a linear fit to the data, but does not have a direct physical meaning. \label{Kzradio}}

\psfig{file=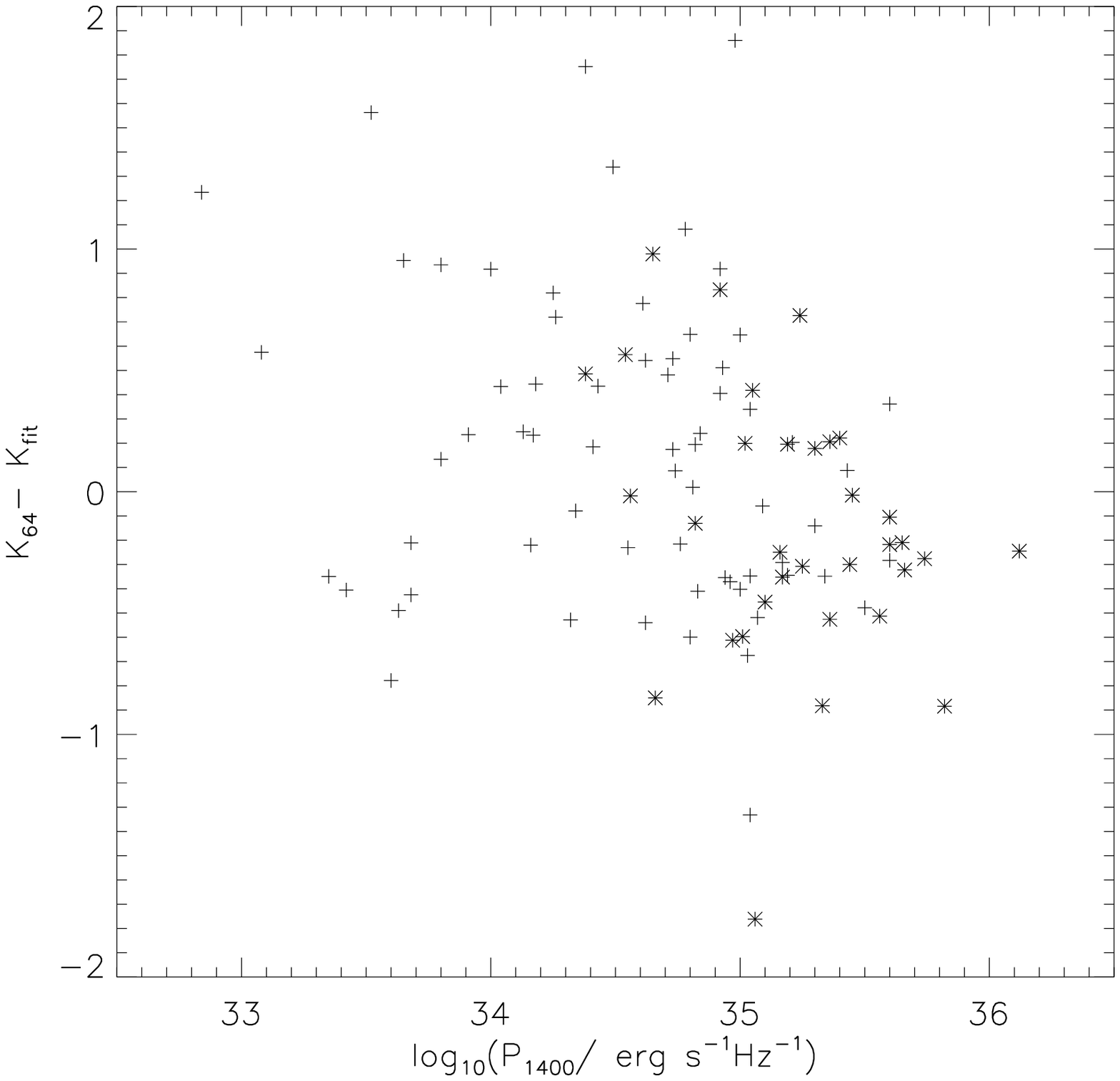,width=16cm}
\figcaption[Kzpower]{Deviation from a linear fit to the $\log_{10}(z) - K_{64kpc}$ relation plotted against radio power at 1.4~GHz, using the data from Figure \ref{Kzradio}. Plus-signs represent radio galaxies at $z<2$, stars those at $z>2$. There is a statistically significant (99.94\%) trend for the higher luminosity radio sources to have more luminous parent galaxies. \label{Kzpower}}

\psfig{file=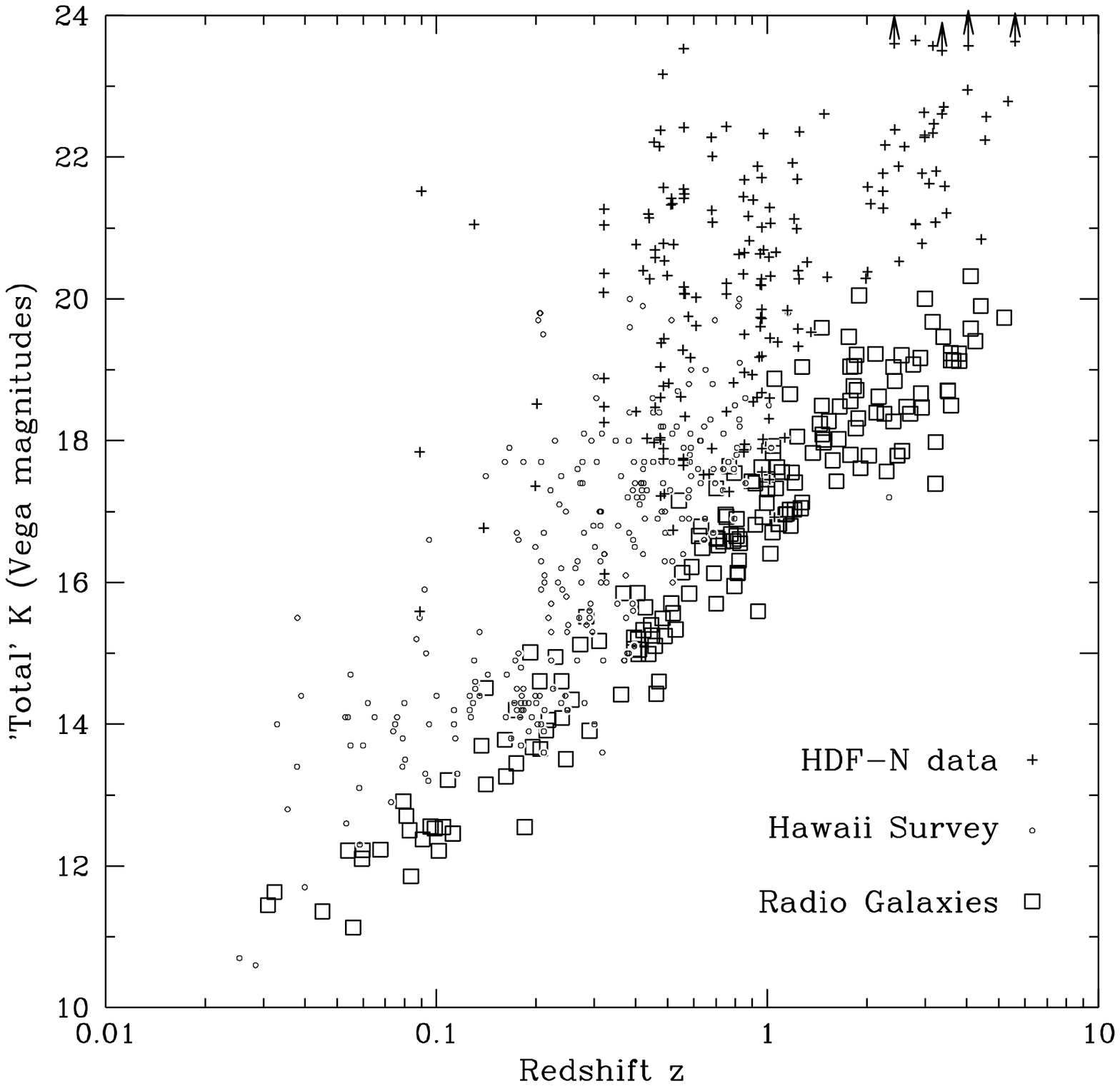,width=16cm}
\figcaption[Kz]{Composite Hubble $K-z$ diagram of radio-loud and radio-quiet galaxies. The radio-loud data are the same as in Figure \ref{Kzradio}; the radio-quiet data are taken from the HDF-North and the Hawaii survey. The radio-quiet galaxies do not show the strong correlation seen in radio-loud galaxies, and are significantly fainter than the radio-loud galaxies, which trace the bright envelope of the radio-quiet data at $z \simlt 1$, and are $\simgt 2$ magnitudes brighter at $z \simgt 1$. \label{Kz}}

\psfig{file=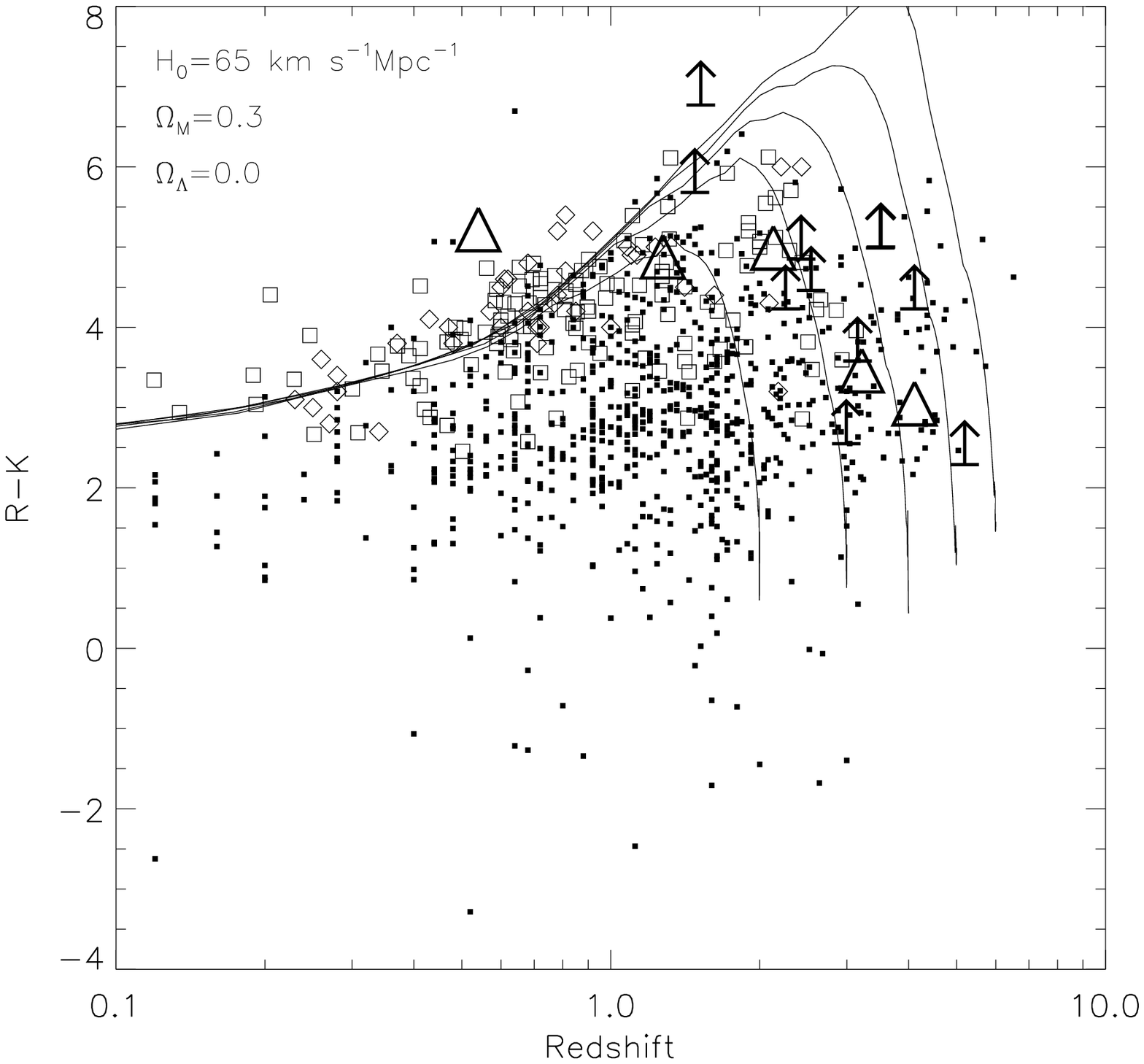,width=16cm}
\figcaption[zRmK]{$R-K$ color plotted against redshift. Small squares are optically selected galaxies from the HDF-North with photometric redshifts \citep{fer99}. Open squares and diamonds are radio galaxies from the MRC sample of \citet{mcc99} and the 7CRS of \citet{wil01b}. Large triangles and lower limits are galaxies from our USS sample. The solid lines show the color evolution of a galaxy formed in an instantaneous burst at $z_{form}=2,3,4,5,6$, calculated using the PEGASE~2 code. \label{zRmK}}

\psfig{file=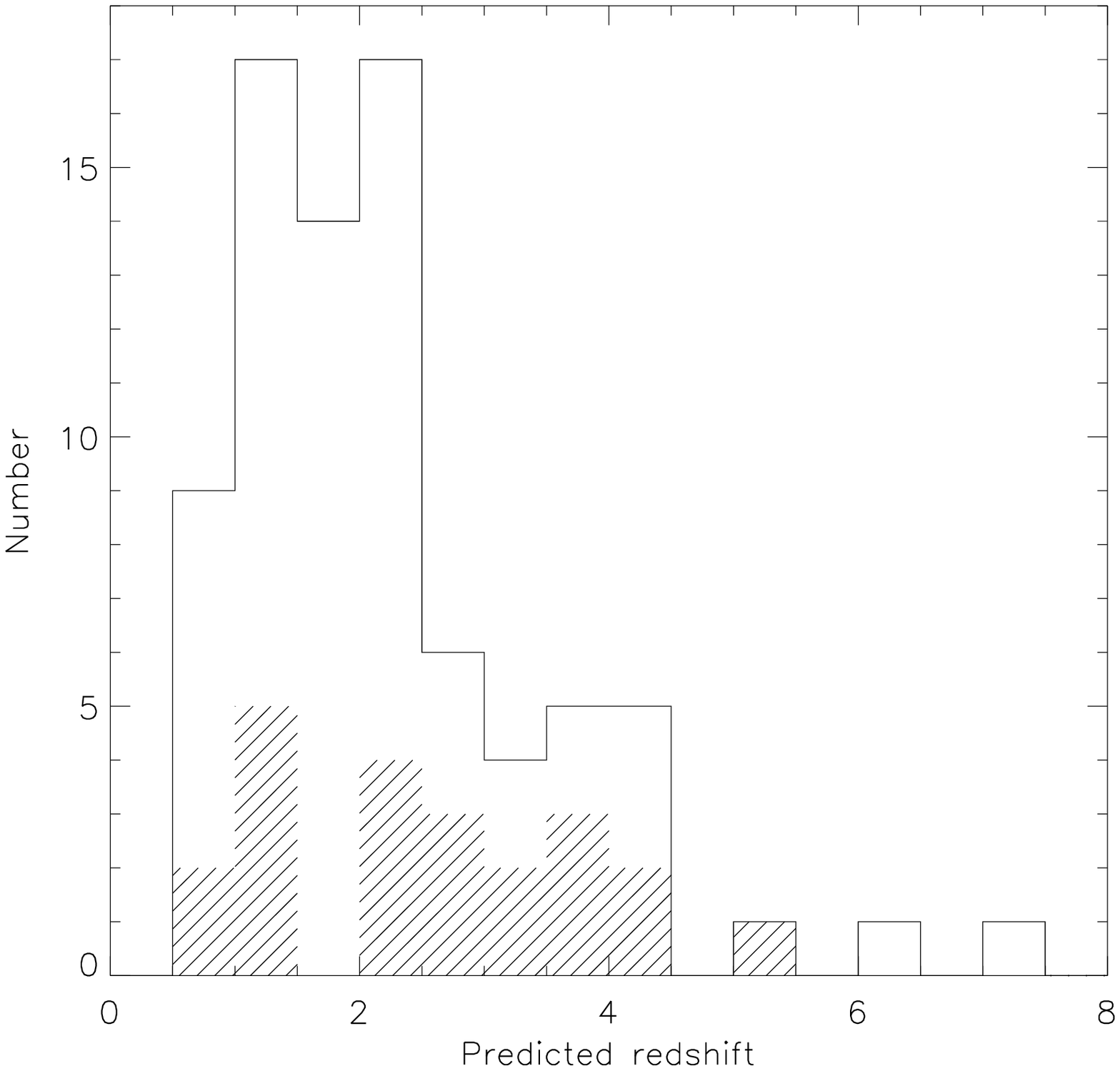,width=16cm}
\figcaption[zpredhis]{Predicted redshift distribution of the 80 objects with $K-$band detections. The redshifts are estimated from the $K-z$ diagram in Figure \ref{Kzradio}. The shaded histogram shows the distribution of the 22 objects with spectroscopic redshifts from this sample. \label{zpredhis}}

\newpage
\appendix

{\bf The individual figures are available on http://www.strw.leidenuniv.nl/$\sim$debreuck/papers/}
\figcaption[fc]{Finding charts for all 128 object observed. We show the near-IR image if obtained, otherwise the optical image, as indicated above each figure. Each image is $90\arcsec \times 90\arcsec$. The circled star labeled 'A' is the offset star, whose coordinates are listed in Table \ref{astrometry}. The position of the radio centroid is indicated with an open crosshair. \label{fc}}
\normalsize

\end{document}